\documentclass[aps,prb,reprint,amsmath,amssymb]{revtex4-2}

\usepackage{amsthm, amsfonts, mathtools}
\usepackage{dsfont, bbold}
\usepackage{braket}
\usepackage{graphicx}
\usepackage{modiagram}
\usepackage{tabularx, adjustbox, multirow}
\usepackage{verbatim}
\usepackage{color}
\usetikzlibrary{positioning}
\usepackage{ulem}
\usepackage[T1]{fontenc}
\usepackage[colorlinks=true,citecolor=blue,linkcolor=blue,urlcolor=blue]{hyperref}

\begin{document}
\title{Predicting Defect States: A Quick Screening Protocol for Substitutional Point Defect Engineering}

\author{Hyosik Kang}
\affiliation{Department of Chemistry, Pennsylvania State University, University Park, Pennsylvania 16802, USA}
\author{Lukas Muechler}
\affiliation{Department of Chemistry, Pennsylvania State University, University Park, Pennsylvania 16802, USA}
\affiliation{Department of Physics, Pennsylvania State University, University Park, Pennsylvania 16802, USA}
\email{lfm5572@psu.edu}

\begin{abstract}
Point defects in crystalline materials play a central role in determining electronic, optical, and magnetic properties. However, systematic exploration of defect configurations remains computationally expensive because large supercell calculations are required to approximate isolated defects under periodic boundary conditions. We present a unit-cell-based tight-binding protocol that enables rapid pre-screening of substitutional defects.
The protocol extracts Wannier tight-binding Hamiltonians from small, fully relaxed unit cells of the host and defect-like systems, replicates the host Hamiltonian to construct a supercell model, and introduces the defect by modifying only the on-site energies at the substitution site while leaving the hopping parameters unchanged.
We validate the protocol across three diverse systems: isostructural substitutional defects in transition-metal dichalcogenides (M$_\mathrm{Mo}$ MoS$_2$, M = Ce, Zr, Nb, Tc, and Ru), symmetry-breaking carbon substitutions in hexagonal boron nitride (C$_\mathrm{B}$C$_\mathrm{N}$ h-BN), and nitrogen-vacancy (NV$^-$) centers in diamond. These case studies span two-dimensional and three-dimensional hosts, simple substitutions, and substitution-vacancy complexes. In all cases, the protocol successfully captures the number of in-gap states, their degeneracies, and their shallow or deep character relative to host band edges, despite some quantitative deviations in absolute energy positions. 
We further identify limitations for vacancies of highly electronegative atoms and for charge-state or spin-polarization effects, both of which involve self-consistent charge redistribution not captured by the protocol. 
\end{abstract}

\date\today
\maketitle

\section{Introduction}
Point defects in crystalline solids arise when the perfect periodic arrangement of atoms is locally disturbed. They can form at finite temperature and appear when foreign atoms are incorporated during synthesis or intentional doping. Because point defects perturb the local bonding environment and electronic structure, they can strongly influence macroscopic material properties. In semiconductors and insulators, many point defects introduce localized defect states in the band gap, which can act as non-radiative recombination centers that limit device efficiency or, conversely, as functional centers that are intentionally engineered. \cite{point_defect_effects_1, point_defect_effects_2, point_defect_effects_3, point_defect_effects_4, point_defect_effects_5, point_defect_effects_6}

Point defects have been identified in crystalline materials as key building blocks for quantum and optoelectronic technologies. \cite{PDforquantum1, PDforquantum2, PDforquantum3, PDforquantum4, PDforquantum5, PDforquantum6} The negatively charged nitrogen-vacancy (NV$^-$) center in diamond is a representative example, combining a localized electronic structure with spin-selective optical transitions and long spin-coherence times, which make it highly attractive for quantum sensing, solid-state single-photon emission, and quantum information applications. \cite{NV-SPE1, NV-SPE2, NV-SPE3, NV-SPE4, NV-SPE5, NV-SPE6, NV-SPE7} In 2D hexagonal boron nitride (h-BN), carbon-related complexes such as the C$_\mathrm{B}$C$_\mathrm{N}$ defect have been proposed as bright and optically stable color centers that can act as single-photon emitters compatible with van der Waals heterostructures, offering a route towards integrated, on-chip quantum light sources. \cite{CBCN_h-BN1, CBCN_h-BN2, CBCN_h-BN3, CBCN_h-BN4, CBCN_h-BN5, CBCN_h-BN6, CBCN_h-BN7} Transition-metal dichalcogenides (TMDCs), such as MoS$_2$ and WS$_2$, are another important class of 2D materials whose electronic and optical properties can be widely tuned by appropriate combinations of transition metals and chalcogen species. \cite{TMDCs1, TMDCs2, TMDCs3, TMDCs4} Among various defect-engineering strategies, lanthanide substitution in TMDCs has attracted interest because $f\text{-}f$ transitions of lanthanide ions typically lie in the infrared spectral range, making such defect centers promising candidates for quantum telecommunication devices. \cite{LnTMDCs1, LnTMDCs2, LnTMDCs3, LnTMDCs4}

Due to their impact on material properties, a variety of theoretical approaches have been developed to model their structure and electronic behavior. A widely adopted strategy is the first-principles supercell approach, in which a defect is introduced into a periodically repeated simulation cell and its properties are computed, typically within density functional theory (DFT). \cite{SupercellApproach1, SupercellApproach2, SupercellApproach3} This framework is highly versatile and can deliver quantitatively accurate formation energies, ground‑state electronic structures, and charge‑transition levels for many classes of defects, especially when combined with embedding methods for excited states. \cite{SupercellApproachQE1, SupercellApproachQE2, SupercellApproachQE3, SupercellApproachQE4} However, each distinct defect configuration generally requires a separate, fully relaxed supercell calculation with sufficiently large cell size and dense k‑point sampling, so the computational cost scales poorly with the number of candidate defects and becomes resource-intensive for systematic high‑throughput screening.

Complementary to fully \textit{ab initio} treatments, tight-binding (TB) models provide a physically intuitive, real-space description of defect states based on localized orbitals and a finite set of on-site and hopping parameters. \cite{TB1, TB2, TB3, TB4} Once a reliable parameterization is available, TB calculations for large supercells and many different defect geometries are computationally inexpensive, making them attractive for studying large supercells and low-dimensional materials where full ab initio treatments are too costly. However, model accuracy is strongly tied to how well the parameters capture local bonding and relaxation around the defect, and extending a given parameter set to many chemically distinct dopants or complex defect complexes is not straightforward. 

More recently, machine‑learning‑based approaches have emerged as powerful tools for accelerating the exploration of point defects. \cite{MLApproach1, MLApproach2, MLApproach3, MLApproach4, MLApproach5} Data‑driven models trained on first‑principles datasets can predict defect structures, formation energies, and selected electronic or optical descriptors across large materials spaces at a fraction of the cost of explicit DFT calculations, thus enabling high-throughput screening of candidate dopants. Nevertheless, machine‑learning approaches typically require substantial training data and careful feature design, and their reliability outside the training domain can be difficult to assess without additional physical validation.

In this context, we propose a protocol for screening defect levels using only simple unit cell calculations, specifically focusing on rapidly identifying promising substitutional defects. Note that the central goal is not to predict exact defect states for every configuration, but to describe key qualitative features such as the number of in-gap states, their degeneracies, and whether the defect states are shallow or deep. This provides a computationally inexpensive way for pre-screening a large number of substitutional candidates and selecting a small subset for more accurate investigation.

The key idea is to approximate the band structure of a defective supercell using only TB Hamiltonians derived from two types of unit cells: the pristine host unit cell and a corresponding “defect‑like” unit cell. The host unit cell is simply the primitive cell of the host material, while the defect‑like unit cell is constructed by replacing the host atom at the relevant lattice site with the substituent species in an otherwise identical crystal structure. For example, in the case of Ce$_\mathrm{Mo}$ MoS$_{2}$, the host unit cell is that of MoS$_2$, whereas the defect‑like unit cell is CeS$_2$ with the same structural symmetry as MoS$_2$. By combining the TB descriptions of these two unit cells in a suitable supercell representation, one can predict the properties, such as the number of expected defect states and their character of a substitutional point defect without performing an explicit defect supercell DFT calculation for each candidate dopant.

\begin{figure}[t]
    \centering
    \includegraphics[width=0.45\textwidth]{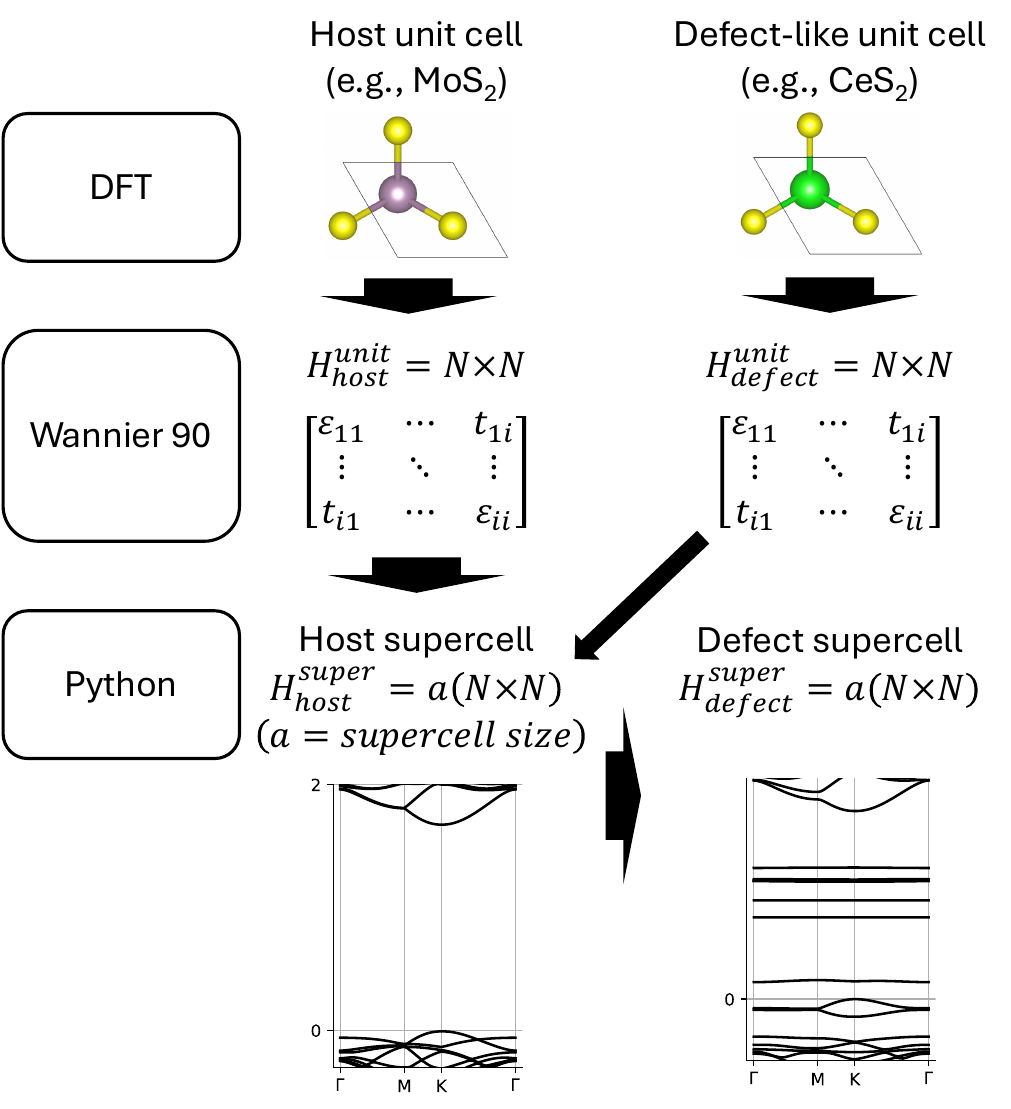}
    \caption{Schematic workflow of the unit-cell-based tight-binding protocol for modeling substitutional defects. Both the host (e.g., MoS$_2$) and the defect-like (e.g., CeS$_2$) unit cells are first fully relaxed using DFT. Localized orbital basis sets are then constructed via Wannier90 to extract the unit cell TB Hamiltonians $H^\mathrm{unit}_\mathrm{host}$ and $H^\mathrm{unit}_\mathrm{defect}$, each of dimension $N \times N$. The host unit cell Hamiltonian is replicated in real space to construct a supercell Hamiltonian $H^\mathrm{super}_\mathrm{host}$ (dimension $aN \times aN$, where $a$ is the number of unit cells in the supercell). A defect is introduced by replacing the on-site parameters at the substitution site with those from $H^\mathrm{unit}_\mathrm{defect}$, yielding the defect supercell Hamiltonian $H^\mathrm{super}_\mathrm{defect}$. Band structures are obtained by diagonalizing the supercell Hamiltonians along selected $\mathbf{k}$-point paths.}
    \label{Workflow}
\end{figure}

The overall workflow, illustrated schematically in Figure~\ref{Workflow}, proceeds as follows. First, both the host and defect‑like unit cells are fully relaxed using DFT, allowing relaxation of lattice vectors and internal atomic coordinates. The rationale for this choice and its impact on TB parameters are discussed in detail in the Results and Discussion section. Second, a localized orbital basis set is constructed for each relaxed unit cell, and the corresponding TB Hamiltonians are extracted. Third, a supercell TB Hamiltonian is generated by replicating the host unit cell Hamiltonian in real space within a Python implementation, thereby constructing a periodic supercell representation of the pristine material. The local Hamiltonian matrix elements associated with the chosen substitutional site are then modified based on the parameters obtained from the defect‑like unit cell, so that the chemical environment of the target site is effectively switched while the surrounding environment remains that of the host. In the present implementation, this substitution primarily modifies on-site energies of the defect site, while leaving the hopping parameters essentially unchanged; the physical rationale for this approximation is discussed in the Results and Discussion section. Finally, the resulting supercell Hamiltonian is diagonalized along the selected k‑points to obtain approximate band structures and the corresponding defect states.

In this paper, we discuss the physical background and rationale of the protocol introduced above and apply it to three representative defect systems: M$_\mathrm{Mo}$ defects in MoS$_2$ (M = Ce, Zr, Nb, Tc, and Ru), the C$_\mathrm{B}$C$_\mathrm{N}$ complex in h‑BN, and the NV$^{-}$ center in diamond. On this basis, we further examine the applicability and limitations of the approach, with particular emphasis on its suitability for substitutional defects and its current non‑spin‑polarized formulation.

\section{Calculation Details}
DFT calculations were performed using the Vienna \textit{ab initio} simulation package (VASP) \cite{Method_VASP} within the generalized-gradient approximation (GGA) in the Perdew--Burke--Ernzerhof (PBE) functional form and the projector augmented-wave (PAW) method. \cite{Method_GGA, Method_PBE, Method_planewavebasisset, Method_PAW} For all compounds, GW-type PAW pseudopotentials were employed to maintain consistency between the host and defect-like unit cells. The plane-wave energy cutoff was set to $500~\text{eV}$, and electronic self-consistency was achieved with an energy convergence threshold of $10^{-6}~\text{eV}$.

All calculations were carried out with full relaxation of both lattice vectors and atomic positions before constructing TB models. For the 2D systems (MoS$_2$ and h-BN), the out-of-plane lattice parameter was chosen to be approximately $10~\text{\AA}$ and was allowed to vary slightly (within about $\pm 0.1~\text{\AA}$) during structural optimization, which is sufficient to avoid spurious interactions under periodic boundary conditions and to capture the intrinsic monolayer behavior. \cite{LnTMDCs4} For Brillouin-zone sampling of the unit cells, Monkhorst--Pack k-point meshes of $5 \times 5 \times 1$ were used for the 2D materials and $8 \times 8 \times 8$ for diamond; these choices represent a compromise between computational cost and the need to include high-symmetry points and obtain well-converged band structures suitable for subsequent Wannier interpolation.

To provide reference data for comparison with the unit-cell-based TB approach, explicit defect supercell calculations were also performed. For the 2D systems, $5 \times 5 \times 1$ supercells were constructed, while a $5 \times 5 \times 5$ supercell was used for diamond. In all supercell calculations, only the $\Gamma$ point was used for Brillouin-zone sampling, as the effect on larger k-meshes to be negligible on band structures. \cite{LnTMDCs4}

TB Hamiltonians were constructed from the converged DFT calculations using the Wannier90 code. \cite{Method_Wannier901} For each system, an initial subspace was defined by projecting the Bloch states onto atom-centered trial orbitals within an appropriate energy window (Mo $d$- and S $p$-orbitals for MoS$_2$, $sp^2$ orbitals for h-BN, and $s$-/$p$-orbitals for diamond). A disentanglement procedure was then applied in all cases to isolate a set of bands corresponding to the chosen number of orbitals, which was used as the basis for the TB Hamiltonian. When needed, we performed further minimization to obtain maximally localized Wannier functions, but in many cases, the disentangled, projected orbitals already provided sufficiently localized functions for constructing the TB Hamiltonian.

For h-BN, it is well established that the conduction band minimum at the $\Gamma$ point has a nearly-free-electron (NFE) character \cite{h-BN_NFE1, h-BN_NFE2} and therefore is not captured by a TB basis built from purely atom-like orbitals. In our projections, the $sp^2$ orbitals appear at high energies from $\Gamma$ to M point. The validation of the resulting Wannier functions and Hamiltonians was assessed by comparing the Wannier-interpolated band structures with the corresponding DFT bands, and by inspecting real-space isosurfaces of the Wannier functions to confirm their orbital character and localization. \cite{Method_Wannier902}

TB calculations were carried out using the Python package PyBinding \cite{Method_PyBinding}, which provides a flexible framework for constructing and diagonalizing lattice Hamiltonians for supercell geometries under periodic boundary conditions. For each material, the TB Hamiltonian of the host unit cell obtained from Wannier90 was used as the building block. A supercell Hamiltonian was generated by replicating this building block in real space to match the supercell sizes used in the reference DFT defect calculations, namely $5 \times 5 \times 1$ for the 2D systems and $5 \times 5 \times 5$ for diamond. Periodic boundary conditions were preserved, ensuring consistency with the corresponding DFT supercell calculations.

To accurately reproduce the DFT and Wannier-interpolated band structures, hopping amplitudes between Wannier orbitals were considered up to the fifth nearest-neighbor shells for the 2D systems (MoS$_2$ and h-BN) and up to the seventeenth nearest-neighbor shells for diamond. Within this supercell Hamiltonian, a substitutional defect was introduced by identifying the target lattice site and modifying the local Hamiltonian matrix elements at that site based on the defect-like unit cell Hamiltonian. The resulting defective supercell Hamiltonians were then diagonalized at each k-point along high-symmetry points to obtain the band structures and defect states for comparison with the explicit DFT supercell calculations.

\section{Results and Discussion}
\subsection{TB background and physical rationale of the unit-cell-based protocol}
Within the TB approach, the electronic structure of a crystalline solid is described in terms of a set of localized orbitals $\{\lvert \phi_{i\alpha} \rangle\}$ centered on lattice sites $i$ and labeled by orbital indices $\alpha$. The Hamiltonian can be written as
\[\hat{H} = 
\sum_{i,\alpha} \varepsilon_{i\alpha}\, \lvert \phi_{i\alpha} \rangle \langle \phi_{i\alpha} \rvert 
+ \sum_{i \neq j} \sum_{\alpha,\beta} t_{i\alpha,j\beta}\, \lvert \phi_{i\alpha} \rangle \langle \phi_{j\beta} \rvert,\]
where $\varepsilon_{i\alpha}$ are on-site energies and $t_{i\alpha,j\beta}$ denote hopping amplitudes between orbitals on different sites. To clarify several types of hopping terms, in monolayer TMDCs, for example, we distinguish hoppings between metal $d$ and chalcogen $p$ orbitals, $t^{\mathrm{M\text{--}X}}_{i\mu,j\nu}$, chalcogen--chalcogen hoppings, $t^{\mathrm{X\text{--}X}}_{i\nu,j\nu'}$, and metal--metal hoppings, $t^{\mathrm{M\text{--}M}}_{i\mu,j\mu'}$. 

Because these TB parameters are defined in a localized, real-space basis, they are sensitive to the local bonding geometry, such as bond lengths and angles around each site. \cite{TB_local1, TB_local2} Conventional calculations of defect states rely on large supercells in which both lattice vectors and internal atomic positions are fully relaxed to obtain the ground-state geometry, and defect states are then computed for this relaxed structure. In our unit-cell-based protocol, a similar idea is implemented at the level of the host and defect-like unit cells: both are fully relaxed so that their bond lengths and angles closely match those in the corresponding explicit defect supercells. 

For the Ce$_\mathrm{Mo}$ MoS$_2$, for instance, we compare the bond lengths in (i) the explicit Ce$_\mathrm{Mo}$ MoS$_2$ supercell, (ii) the host MoS$_2$ unit cell, (iii) the fully relaxed defect-like CeS$_2$ unit cell, and (iv) the defect-like CeS$_2$ unit cell with lattice vectors fixed to those of the host in Table~\ref{SI_Table1}. For the host MoS$_2$ unit cell, the agreement with the local geometry around the Mo site in the explicit Ce$_\mathrm{Mo}$ MoS$_2$ supercell is already excellent. Consequently, the on-site energies at Mo and the Mo--S hopping amplitudes are nearly identical between the two. For the defect-like CeS$_2$ unit cell, although both configurations show some deviations in Ce--S bond lengths and angles relative to the explicit Ce$_\mathrm{Mo}$ MoS$_2$ supercell, the fully relaxed CeS$_2$ unit cell reproduces the local bonding environment around Ce noticeably better, and the corresponding TB on-site energies at Ce and Ce--S hopping amplitudes align more closely with those extracted from the explicit defect supercell. This difference is particularly pronounced for the Ce $f$-orbital on-site energies: the fully relaxed CeS$_2$ unit cell yields $f$-level positions that differ by less than $\sim 0.12$ eV from the explicit defect system, whereas the lattice-fixed CeS$_2$ unit cell shows deviations of approximately 1 eV. Such a discrepancy would shift the predicted position of the Ce $f$ defect levels substantially and compromise the reliability of the screening protocol. Consequently, we adopt the fully relaxed defect-like unit cells throughout this work.

The real-space localization of the Wannier orbitals also implies a rapid decay of the hopping amplitudes $t_{i\alpha,j\beta}$ with increasing distance between sites $i$ and $j$. \cite{Method_Wannier901, Method_Wannier902, TB_local1, TB_local2} Similarly, point defects are generally localized and therefore primarily controlled by on-site energies and hopping amplitudes within the first nearest-neighbor shell around the defect. In MoS$_2$ and CeS$_2$, representative metal-metal hopping amplitudes ($t^{\mathrm{M\text{--}M}}_{i\mu,j\mu}$) decrease by roughly an order of magnitude between the first and second nearest-neighbor shells, and higher-shell contributions become progressively less important (see Fig.~\ref{SI_Fig_NN}). This behavior provides the physical basis for our unit-cell-based protocol: by using TB parameters from relaxed host unit cells and modifying parameters only on the substitutional site from the parameters obtained from the defect‑like unit cell, one can capture the main features of the defect states in a large supercell without performing an explicit defect supercell DFT calculation for each defect configuration.

\subsection{Case study 1: \texorpdfstring{M$_\mathrm{Mo}$ MoS$_{2}$\\ (M = Ce, Zr, Nb, Tc, and Ru)}{M(Mo) MoS2 (M = Ce, Zr, Nb, Tc, Ru)}}
\begin{table*}[!htb]
    \centering
    \renewcommand{\arraystretch}{1.3}
    \setlength{\tabcolsep}{6pt}
    \begin{tabular}{cc|c|ccccc}
    \multirow{2}{*}{Terms} & \multirow{2}{*}{Symmetry}
          & CeS$_2$ & ZrS$_2$  & NbS$_2$  & MoS$_2$  & TcS$_2$  & RuS$_2$  \\
        & 
          & (Ce$^{4+}$, 6d$^{0}$) & (Zr$^{4+}$, 5d$^{0}$) & (Nb$^{4+}$, 5d$^{1}$) & (Mo$^{4+}$, 5d$^{2}$) & (Tc$^{4+}$, 5d$^{3}$) & (Ru$^{4+}$, 5d$^{4}$)  \\[0.1em]
    \hline
    \multirow{6}{*}{$\varepsilon^\mathrm{M}$} 
          &A$'_1$                 &3.65   &2.07   &0.78   &-0.19  &-2.04  &-2.32 \\
          &($d_{z^{2}}$)          &(+3.84)&(+2.26)&(+0.97)&-      &(-1.85)&(-2.13)\\[0.2em]
          &E$'$                   &3.65   &1.97   &0.81   &-0.06  &-1.97  &-2.17\\
          &($d_{x^2-y^2},d_{xy}$) &(+3.71)&(+2.03)&(+0.87)&-      &(-1.91)&(-2.11)\\[0.2em]
          &E$''$                  &4.55   &2.96   &1.83   &0.94   &-1.09  &-1.50 \\
          &($d_{xz},d_{yz}$)      &(+3.61)&(+2.02)&(+0.89)&-      &(-2.03)&(-2.44)\\[0.2em]
    \multirow{4}{*}{$\varepsilon^\mathrm{X}$}
          &A$_1$                  &-0.34  &-0.85  &-1.55  &-1.94  &-3.23  &-2.85  \\
          &($p_z$)                &(+1.6) &(+1.09)&(+0.39)&-      &(-1.29)&(-0.91)\\[0.2em]
          &E                        &-0.26  &-1.00  &-1.52  &-1.78  &-3.40  &-3.10  \\
          &($p_x,p_y$)            &(+1.52)&(+0.78)&(+0.26)&-      &(-1.62)&(1.32) \\[0.2em]
    \hline
    \multirow{6}{*}{$t^{\mathrm{M\text{--}X}}_{i\mu,j\nu}$}
        & \multirow{2}{*}{A$'_1$--A$_1$}
                                    &1.85   &1.78   &1.81   &1.78   &1.80   &1.69   \\
        &                           &(+0.07)&(+0)   &(+0.03)&-      &(+0.02)&(-0.09)\\
        & \multirow{2}{*}{E$'$--E}
                                    &2.11   &1.91   &1.93   &1.92   &2.11   &2.08   \\
        &                           &(+0.19)&(-0.01)&(+0.01)&-      &(+0.19)&(+0.16)\\
        & \multirow{2}{*}{E$''$--E}
                                    &1.44   &1.38   &1.45   &1.48   &1.55   &1.52   \\
        &                           &(-0.04)&(-0.1) &(-0.03)&-      &(+0.07)&(+0.04)\\
    \multirow{4}{*}{$t^{\mathrm{X_{up}\text{--}X_{dn}}}_{i\nu,j\nu}$}
        & \multirow{2}{*}{A--A}
                                    &0.83   &1.07   &1.08   &1.10   &1.37   &1.49   \\
        &                           &(-0.27)&(-0.03)&(-0.02)&-      &(+0.27)&(+0.39)\\
        & \multirow{2}{*}{E--E}
                                    &0.16   &0.06   &0.06   &0.06   &0.01   &0.08   \\
        &                           &(+0.1) &(+0)   &(+0)   &-      &(-0.05)&(+0.02)\\
    \hline
    \multirow{6}{*}{$t^{\mathrm{M\text{--}M}}_{i\mu,j\mu}$}
        & \multirow{2}{*}{A$'_1$--A$'_1$}
                                    &0.16  &0.19   &0.25   &0.29   &0.16   &0.12   \\
        &                           &(-0.13)&(-0.1) &(-0.04)&-      &(-0.13)&(-0.17)\\
        & \multirow{2}{*}{E$'$--E$'$}
                                    &0.19  &0.12   &0.13   &0.14   &0.12   &0.10   \\
        &                           &(+0.05)&(-0.02)&(-0.01)&-      &(-0.02)&(-0.04)\\
        & \multirow{2}{*}{E$''$--E$''$}
                                    &0.13  &0.11   &0.10   &0.10   &0.13   &0.13   \\
        &                           &(+0.03)&(+0.01)&(+0)   &-      &(+0.03)&(+0.03)\\
    \multirow{4}{*}{$t^{\mathrm{X\text{--}X}}_{i\nu,j\nu}$}
        & \multirow{2}{*}{A$_1$--A$_1$}
                                    &0.06  &0.11   &0.15   &0.21   &0.16   &0.16   \\
        &                           &(-0.15)&(-0.1) &(-0.06)&-      &(-0.05)&(-0.05)\\
        & \multirow{2}{*}{E--E}
                                    &1.00   &0.24   &0.32   &0.41   &0.37   &0.37   \\
        &                           &(+0.59)&(-0.17)&(-0.09)&-      &(-0.04)&(-0.04)\\
    \hline
    \end{tabular}       
    \caption{The first block ($\varepsilon^\mathrm{M}$ and $\varepsilon^\mathrm{X}$) is on-site energies relative to the Fermi level of each system, i.e., as $E - E_F$. The second block ($t^{\mathrm{M\text{--}X}}_{i\mu,j\nu}$ and $t^{\mathrm{X_{up}\text{--}X_{dn}}}_{i\nu,j\nu}$) is hopping amplitudes between different orbitals, listed as absolute values and restricted to terms with magnitudes exceeding 0.01. The third block ($t^{\mathrm{M\text{--}M}}_{i\mu,j\mu}$ and $t^{\mathrm{X\text{--}X}}_{i\nu,j\nu}$) is hopping amplitudes between same orbitals but different site, considering only the first nearest-neighbor shell; these values are averaged to account for direction-dependent variations under symmetry. Each orbital is grouped by local symmetry (A$'$$_1$, E$'$, E$''$, etc.). The values in parentheses indicate the deviation from the corresponding MoS$_2$ unit cell Hamiltonian values.}
    \label{Case study 1}
\end{table*}

\begin{figure*}[t]
    \centering
    \includegraphics[width=0.95\textwidth]{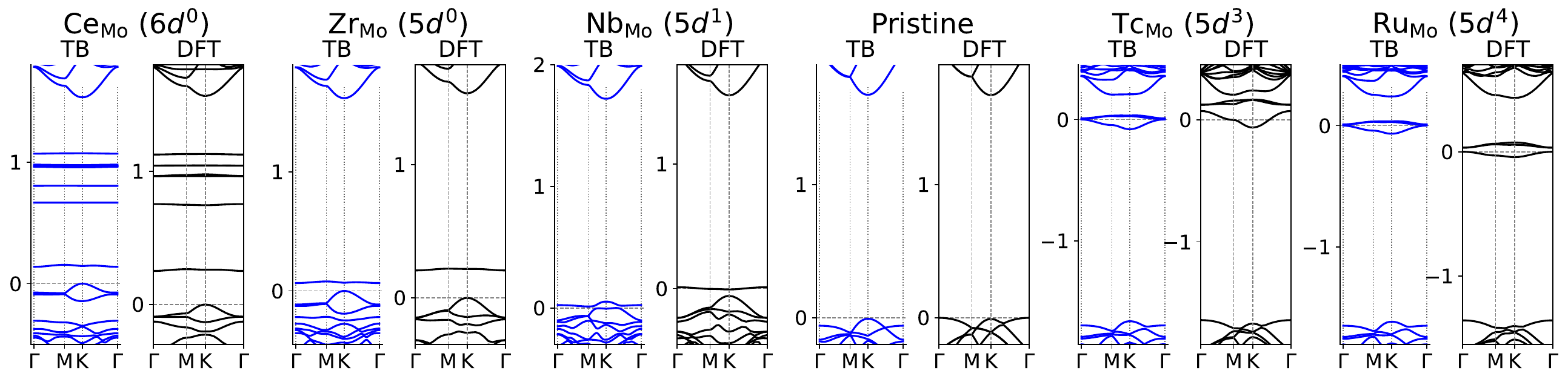}
    \caption{Band structures of Ce$_\mathrm{Mo}$, Zr$_\mathrm{Mo}$, Nb$_\mathrm{Mo}$, Pristine MoS$_2$, Tc$_\mathrm{Mo}$, and Ru$_\mathrm{Mo}$. The left (blue) bands are obtained from the unit-cell-based protocol, and the right (black) bands are calculated from DFT supercell calculations. The Fermi energy is set at $E_F = 0\ eV$.}
    \label{Fig_Case1_BS}
\end{figure*}

Point defects in Monolayer MoS$_2$ and WS$_2$ have been widely studied, as an atomically thin geometry facilitates the identification of point defect composition and structure, and enables direct measurement of defect-derived electronic and optical properties using local probes. \cite{TMDCs1, TMDCs2, TMDCs3, TMDCs4}  This makes them appealing for systematically exploring how substitutional point defects modify band structures and introduce localized states.\cite{Case_study_1_TMDCs2}

Among the various point defects explored in MoS$_2$ and WS$_2$, lanthanide substitutional defects have recently attracted attention. The main reason is that most lanthanide ions exhibit sharp intra $f\text{-}f$ optical transitions in the infrared spectral range, suggesting the potential as single-photon emitters operating at quantum telecommunication. 

However, recent theoretical studies on Ce$_\mathrm{Mo}$ and Er$_\mathrm{Mo}$ MoS$_2$ and WS$_2$  have shown that, in addition to the desired $f$-orbital states, additional defect states appear in the band gap. \cite{LnTMDCs2, LnTMDCs4} These extra states act as potential noise for $f\text{-}f$ optical transitions and originate from the mismatch in $d$-electron configurations between the host transition metal ion (Mo$^{4+}$ or W$^{4+}$) and the substituting lanthanide ion (Ce$^{4+}$ or Er$^{3+}$), which produces a substantial local potential difference at the substitutional site. Moreover, they have shown that Zr$_\mathrm{Mo}$ MoS$_2$ also exhibits the same defect states, in which Zr$^{4+}$ has the same $d^0$ configuration as Ce$^{4+}$ and Er$^{3+}$. This observation motivates the use of Ce$_\mathrm{Mo}$ as the most direct $f$-orbital-containing test case, together with a series of simpler $5d$ transition metal atoms to assess the performance of the proposed unit-cell-based protocol.

In the remainder of this subsection, the protocol is therefore applied to substitutional defects in MoS$_2$, namely Ce$_\mathrm{Mo}$, Nb$_\mathrm{Mo}$, Tc$_\mathrm{Mo}$, Ru$_\mathrm{Mo}$, and Zr$_\mathrm{Mo}$, corresponding to different $d$-electron configurations from $d^0$ to $d^4$ for M$^{4+}$ ion. Tc and Ru are included here as systematic benchmark cases to evaluate how well the protocol reproduces defect states across a controlled sequence of substitutional environments. For Ce$_\mathrm{Mo}$, a 19-orbital basis is employed to explicitly include the $f$-derived states of the defect-like CeS$_2$ unit cell; specifically, this basis contains the 11 orbitals used for the MoS$_2$ host cell plus 7 $f$-orbitals and 1 vacuum state, that lie within the band gap of the defect-like unit cell. For the remaining cases, the TB model is constructed in the same 11-orbital basis (5 metal $d$- and 6 chalcogen $p$-orbitals) for both the host and defect-like unit cells, ensuring a one-to-one correspondence of basis functions. This setup allows a direct comparison between the defect states predicted by the unit-cell-based protocol and those obtained from explicit DFT supercell calculations.

To modify the Hamiltonian elements associated with the defect site in the host supercell Hamiltonian, we compare the unit-cell Hamiltonians obtained for each system in Table~\ref{Case study 1}. The 11 orbitals can be grouped by local symmetry into A$'$$_1$, E$'$, E$''$, A$_1$, and E. The off-diagonal terms, i.e., hopping amplitudes between different orbitals, corresponding to $t^{\mathrm{M\text{--}X}}_{i\mu,j\nu}$ and $t^{\mathrm{X_{up}\text{--}X_{dn}}}_{i\nu,j\nu}$, are nonzero only for orbitals with matching symmetry character. For same-orbital hopping between different sites, $t^{\mathrm{M\text{--}M}}_{i\mu,j\mu}$ and $t^{\mathrm{X\text{--}X}}_{i\nu,j\nu}$, only the first nearest-neighbor shell is considered to probe the local-environment change induced by the point defect, and the values are averaged to account for direction-dependent variations under symmetry. To facilitate comparison of the orbital overlap across different systems without the complication of sign changes arising from symmetry and directional conventions, all hopping amplitudes in Table~\ref{Case study 1} are presented in absolute values. The values in parentheses indicate the deviation from the corresponding MoS$_2$ unit-cell Hamiltonian values.

The on-site energies in the TB Hamiltonians are aligned with respect to the Fermi level of each unit cell, and we therefore work with $E - E_F$ values. This choice reflects the fact that, in monolayer TMDCs such as MoS$_2$, the band gap is largely controlled by the crystal-field splitting and hybridization of the transition-metal $d$-orbitals with the chalcogen $p$-orbitals. \cite{Case_study_1_TMDCs1, Case_study_1_TMDCs2} In such systems, the $d$-derived states participate actively in screening and in setting the position of the Fermi level. As the $d$-electron count of the substituting ion is varied, these screening and local-bonding effects readjust the relative alignment between the metal $d$ and chalcogen $p$ manifolds. Aligning on-site energies with respect to the Fermi level of each unit cell and comparing on-site shifts across the substitutional series, therefore, provides a natural way to capture changes in the local chemical environment around the defect site.

For CeS$_2$, the Ce$^{4+}$ ion nominally has the same $d$-electron configuration as Zr$^{4+}$, but its $6d$ shell and correspondingly larger ionic radius compared to 5$d$ transition metals lead to differences in the TB Hamiltonian parameters compared with ZrS$_2$ (Table~\ref{Case study 1}). As a result, the deviations of the hopping elements from the MoS$_2$ values (numbers in parentheses) are enhanced: they range approximately from $-0.27$ eV $(t^{\mathrm{X_{up}\text{--}X_{dn}}}_{i\nu,j\nu},\,\text{A--A})$ up to $+0.59$ eV $(t^{\mathrm{X\text{--}X}}_{i\nu,j\nu},\,\text{E--E})$, which is larger than the corresponding range for ZrS$_2$ $\bigl($from $-0.17$ eV $(t^{\mathrm{X\text{--}X}}_{i\nu,j\nu},\,\text{E--E})$ to $+0.01$ eV $(t^{\mathrm{M\text{--}M}}_{i\nu,j\nu},\,\text{E}''\text{--E}'')\bigr)$. However, these hopping variations are roughly an order of magnitude smaller than the on-site energy shifts at the Ce site, which span from $+2.71$ eV $(\varepsilon^\mathrm{M},\,\text{E}'')$ to $+4.61$ eV $(\varepsilon^\mathrm{M},\,\text{E}')$.

This qualitative behavior of large on-site energies compared to hoppings is not unique to Ce$_\mathrm{Mo}$ but holds more generally across the 5$d$ transition-metal substitutional series from ZrS$_2$ to RuS$_2$ (Table~\ref{Case study 1}). The hopping terms, measured as deviations from the MoS$_2$ host unit-cell Hamiltonian (values in parentheses), typically range from about $-0.17$ eV $(t^{\mathrm{X\text{--}X}}_{i\nu,j\nu},\,\text{E--E for ZrS}_2)$ to $+0.39$ eV $(t^{\mathrm{X_{up}\text{--}X_{dn}}}_{i\nu,j\nu},\,\text{A--A for RuS}_2)$, with most entries having absolute values smaller than 0.2 eV. By contrast, the on-site energy shifts span a much wider range, from $-2.44$ eV $(\varepsilon^\mathrm{M},\,\text{E}'' \text{ for RuS}_2)$ up to $+2.26$ eV $(\varepsilon^\mathrm{M},\,\text{A}'$$_1$ for ZrS$_2)$, i.e., roughly an order of magnitude larger than the typical hopping variations.

This behavior indicates that, despite small differences in bond lengths and angles, the overall strength of orbital hybridization between atoms remains relatively similar across the substitutional series as long as the global crystal structure and local symmetry are preserved. For the present unit-cell-based protocol, this has an important implication: since the substitutional defect does not strongly break the global symmetry of the lattice, the influence of changes in hopping amplitudes on the character of the defect states is relatively small, whereas the shifts in on-site energies play the dominant role in determining the position and nature of the defect states.

Consequently, when implementing the protocol, rather than modifying all Hamiltonian matrix elements around the defect, we restrict the modification to the on-site terms at the substituted atom site in the supercell Hamiltonian. In practice, for the Ce$_\mathrm{Mo}$, Zr$_\mathrm{Mo}$, Nb$_\mathrm{Mo}$, Tc$_\mathrm{Mo}$, and Ru$_\mathrm{Mo}$ defects in MoS$_2$, we start from the host MoS$_2$ supercell TB Hamiltonian and replace the Mo on-site energies $\varepsilon^\mathrm{M}$ at the substitution site, $-0.19$ eV (A$'$$_1$), $-0.06$ eV (E$')$, and $0.94$ eV (E$''$), with the corresponding defect-like unit-cell values: $3.65$ eV (A$'$$_1$), $3.65$ eV (E$')$, and $4.55$ eV (E$''$) for CeS$_2$; $2.07$ eV (A$'$$_1$), $1.97$ eV (E$')$, and $2.96$ eV (E$''$) for ZrS$_2$; $0.78$ eV (A$'$$_1$), $0.81$ eV (E$')$, and $1.83$ eV (E$''$) for NbS$_2$; $-2.04$ eV (A$'$$_1$), $-1.97$ eV (E$')$, and $-1.09$ eV (E$''$) for TcS$_2$; and $-2.32$ eV (A$'$$_1$), $-2.17$ eV (E$')$, and $-1.50$ eV (E$''$) for RuS$_2$.

In addition, for Ce$_\mathrm{Mo}$, to describe the Ce$^{4+}$-derived $f$-like defect states inside the band gap, the original 275$\times$275 MoS$_2$ supercell TB Hamiltonian is augmented by the seven $f$-orbital degrees of freedom obtained from the CeS$_2$ defect-like unit-cell TB Hamiltonian (Table~\ref{SI_Table1}), resulting in a 282$\times$282 Hamiltonian. In this construction, not only the on-site energies of the $f$ orbitals but also their associated hoppings are included, since these hoppings are absent in the MoS$_2$ host unit cell Hamiltonian.

Even with this minimal change, the band structures obtained from the TB supercell (shown in blue in Fig.~\ref{Fig_Case1_BS}) agree well with the reference DFT supercell band structures (shown in black in Fig.~\ref{Fig_Case1_BS}) in terms of the key qualitative features: the number of in-gap states, their degeneracies, and whether the defect levels are shallow or deep relative to the host band edges. While the absolute energy positions of the defect levels may show some quantitative deviations up to $\sim 0.2$ eV in this system, the essential characteristics relevant for defect screening are captured reliably. Since the central goal of the protocol is to provide a computationally inexpensive way to pre-screen a large number of substitutional candidates and select only a small subset for more accurate calculations, the following case studies likewise focus on band structures obtained by modifying only the on-site energies of the substituted atoms in the supercell Hamiltonians.

\subsection{Case study 2: C\texorpdfstring{$_\mathrm{B}$}{B}C\texorpdfstring{$_\mathrm{N}$}{N} h-BN}
\begin{table*}[!htb]
    \centering
    \renewcommand{\arraystretch}{1.3}
    \setlength{\tabcolsep}{6pt}
    \begin{tabular}{cc|ccc}
    \multirow{2}{*}{Terms} & \multirow{2}{*}{Symmetry}
                        &BN             &BC$_\mathrm{N}$  &C$_\mathrm{B}$N  \\
    &                   &(h-BN)         &(C substitutes N)  &(C substitutes B)  \\[0.1em]
    \hline
    \multirow{4}{*}{$\varepsilon^{\alpha}$} 
      &\multirow{2}{*}{$sp^2$}
                        &8.2            &4.87               &4.37               \\
      &                 &               &(-3.13)            &(-3.83)            \\[0.2em]
      &\multirow{2}{*}{$p_z$}
                        &4.79           &2.9                &3.28               \\
      &                 &               &(-1.89)            &(-1.51)            \\[0.2em]
    \multirow{4}{*}{$\varepsilon^{\beta}$} 
      &\multirow{2}{*}{$sp^2$}
                        &-1.4           &0.44               &-1.26              \\
      &                 &               &(+1.84)            &(+0.14)            \\[0.2em]
      &\multirow{2}{*}{$p_z$}
                        &-2.14          &0.34               &0.41               \\
      &                 &               &(+2.48)            &(+2.55)            \\[0.2em]
    \hline
    \multirow{6}{*}{$t^{\alpha\text{--}\beta}_{i\mu,j\nu}$}
        & \multirow{2}{*}{$sp^2$--$sp^2$}
                        &3.99           &4.02               &5.08   \\
        &               &               &(+0.03)            &(+1.09)\\
        & \multirow{2}{*}{$sp^2$--$p_z$}
                        &1              &0                  &0      \\
        &               &               &(-1)               &(-1)   \\
        & \multirow{2}{*}{$p_z$--$p_z$}
                        &9.77           &8.14               &9.21   \\
        &               &               &(-1.63)            &(-0.56)\\
    \hline
    \multirow{4}{*}{$t^{\alpha\text{--}\alpha}_{i\mu,j\mu}$}
      &\multirow{2}{*}{$sp^2$--$sp^2$}
                        &0.08           &0.04               &0.17               \\
      &                 &               &(-0.04)            &(+0.09)            \\[0.2em]
      &\multirow{2}{*}{$p_z$--$p_z$}
                        &0.01           &0.18               &0.17                \\
      &                 &               &(+0.17)            &(+0.16)             \\[0.2em]
    \multirow{4}{*}{$t^{\beta\text{--}\beta}_{i\mu,j\mu}$}
      &\multirow{2}{*}{$sp^2$--$sp^2$}
                        &0.16           &0.2                &0.25               \\
      &                 &               &(+0.04)            &(+0.09)            \\[0.2em]
      &\multirow{2}{*}{$p_z$--$p_z$}
                        &0.23           &0.17               &0.17                \\
      &                 &               &(-0.06)            &(-0.06)            \\[0.2em]
    \hline
    \end{tabular}       
    \caption{The first block lists the on-site energies $\varepsilon^\alpha$ and $\varepsilon^\beta$, where $\alpha$ denotes the first atomic site (B in h-BN, B in C$_\mathrm{B}$N, C in CN$_\mathrm{B}$) and $\beta$ denotes the second atomic site (N in h-BN, C in C$_\mathrm{B}$N, N in CN$_\mathrm{B}$). Each atomic site is described by three $sp^2$ hybrid orbitals and one $p_z$ orbital. All on-site energies are aligned with respect to the valence-band maximum (VBM) of the pristine h-BN unit cell. The second block ($t^{\alpha\text{--}\beta}_{i\mu,j\nu}$) presents hopping amplitudes between different orbitals, listed as absolute values. The third block ($t^{\alpha\text{--}\alpha}_{i\mu,j\mu}$ and $t^{\beta\text{--}\beta}_{i\nu,j\nu}$) presents hopping amplitudes between same orbitals at different sites, considering only the first nearest-neighbor shell; these values are averaged to account for direction-dependent variations under symmetry. The values in parentheses indicate the deviation from the corresponding h-BN unit-cell Hamiltonian values.}
    \label{Case study 2}
\end{table*}

\begin{figure}[t]
    \centering
    \includegraphics[width=0.2\textwidth]{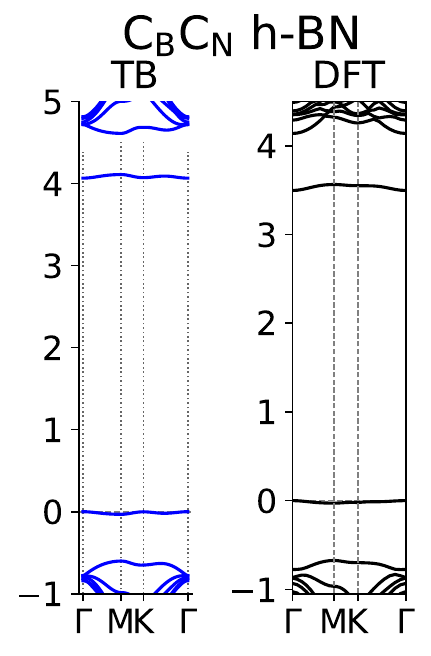}
    \caption{Band structures of C$_\mathrm{B}$C$_\mathrm{N}$ h-BN. The left (blue) bands are obtained from the unit-cell-based protocol, and the right (black) bands are calculated from DFT supercell calculations. The Fermi energy is set at $E_F = 0\ eV$.}
    \label{Fig_Case2_BS}
\end{figure}

In hexagonal boron nitride (h-BN), the C$_\mathrm{B}$C$_\mathrm{N}$ defect, where carbon atoms substitute adjacent boron and nitrogen sites, introduces localized $\pi$-type deep defect states within the wide band gap. This carbon dimer defect exhibits a zero-phonon line at approximately 4.1 eV, making it a bright single-photon emitter that is stable at room temperature. The defect can be integrated into van der Waals heterostructures, making it a practical candidate for on-chip quantum photonics applications. \cite{CBCN_h-BN1, CBCN_h-BN2, CBCN_h-BN3, CBCN_h-BN4, CBCN_h-BN5, CBCN_h-BN6, CBCN_h-BN7}

Structurally, the C$_\mathrm{B}$C$_\mathrm{N}$ defect preserves the local honeycomb lattice geometry of h-BN, with carbon atoms occupying nearest-neighbor B and N sublattice sites. This configuration introduces a pair of carbon atoms into an otherwise more ionic B--N lattice, resulting in a local bonding environment that more closely resembles graphene. Also, this leads to the local $C_\mathrm{3v}$ symmetry of the defect complex, which breaks the inversion symmetry of the pristine h-BN lattice.

The C$_\mathrm{B}$C$_\mathrm{N}$ h-BN system serves as an important second case study for the unit-cell-based protocol because it represents a distinct scenario from the TMDC substitutional defects examined earlier. In the TMDC cases, both the global crystal structure and the local point-group symmetry around the defect site are preserved upon substitution. In contrast, the C$_\mathrm{B}$C$_\mathrm{N}$ defect maintains the global honeycomb lattice but locally breaks inversion symmetry. This difference tests whether the protocol can capture defect states even when local symmetry is reduced.

To model the C$_\mathrm{B}$C$_\mathrm{N}$ defect, we compute TB Hamiltonians for the host h-BN unit cell and two defect-like unit cells with the same honeycomb lattice structure: BC$_\mathrm{N}$ (where carbon substitutes nitrogen), and C$_\mathrm{B}$N (where carbon substitutes boron). These defect-like unit cells are chosen to capture the changes in TB parameters arising from the distinct C--N, and C--B bonding environments introduced by the C$_\mathrm{B}$C$_\mathrm{N}$ defect complex. To maintain a consistent orbital basis across all systems, each atomic site in every unit cell is described by three $sp^2$ hybrids and one $p_z$ orbital. The TB Hamiltonian parameters extracted from these unit cells are summarized in Table~\ref{Case study 2}. Same as TMDC, all hopping amplitudes in Table~\ref{Case study 2} are presented in absolute values. Also, the same-orbital hopping between different sites, $t^{\alpha\text{--}\alpha}_{i\mu,j\mu}$, only the first nearest-neighbor shell is considered, and the values are averaged to account for direction-dependent variations under symmetry. The values in parentheses indicate the deviation from the corresponding h-BN unit cell Hamiltonian values.

The on-site energies in the TB Hamiltonians in this case, however, require a different alignment strategy than the TMDC case. In wide-gap $sp$-bonded materials such as h-BN, the separation between valence and conduction band is comparatively rigid against changes in the local defect configuration. If the on-site energies of the defect-like unit cell were aligned directly to its Fermi level, the additional electrons introduced by the defect would pin $E_\mathrm{F}$ near the host conduction-band minimum (CBM), leading to an artificial downward shift of the defect levels and an underestimation of their positions within the gap. To avoid this artifact, we instead align the defect-like unit cell on-site energies to the valence-band maximum (VBM) of the corresponding pristine host before constructing the TB supercell models. This VBM-based alignment provides a common and physically meaningful reference, allowing for the accurate comparison of defect levels relative to the band edges of the wide-gap host.

From the TB Hamiltonian data in Table~\ref{Case study 2}, the deviations in both hopping amplitudes and on-site energies are larger than those observed in the TMDC substitutional series. The hopping terms, measured as deviations from the h-BN host unit cell Hamiltonian (values in parentheses), range from approximately $-1.63$ eV ($t^{B\text{--}C}_{i\mu,j\nu}$, $p_z$--$p_z$) to $+1.09$ eV ($t^{C\text{--}N}_{i\mu,j\nu}$, $sp^2$--$sp^2$), while the on-site energy shifts span from $-3.83$ eV ($\varepsilon^{C}$, $sp^2$) to $+2.55$ eV ($\varepsilon^{N}$, $p_z$). Unlike the TMDC cases, where on-site variations were roughly an order of magnitude larger than hopping variations, the two quantities are more comparable in magnitude for the h-BN system.

This behavior reflects the distinct bonding character of h-BN compared to the defect-like unit cells. In pristine h-BN, the B--N bond exhibits partial ionic character due to the electronegativity difference between boron and nitrogen, leading to significant charge localization on the nitrogen sites and a correspondingly large band gap. When carbon substitutes either boron or nitrogen, the local bonding environment shifts toward a more covalent character, and the valence electrons become relatively delocalized compared to the ionic B--N lattice. This delocalization leads to larger changes in the hopping amplitudes. Furthermore, the introduction of carbon atoms with different valence electron occupancy induces charge redistribution and distortions of the ideal $sp^2$ hybridization, which further modify the hopping parameters.

Despite these larger hopping variations, the on-site energies remain the dominant factor in determining the positions of the defect states. Recalling that the central goal of the protocol is to provide a computationally inexpensive way to capture the key qualitative features of defect states, when implementing the protocol for the C$_\mathrm{B}$C$_\mathrm{N}$ defect, we again restrict the modification to the on-site terms at the substituted atom sites in the supercell Hamiltonian, leaving the hopping parameters unchanged from the h-BN host values. In practice, we start from the host h-BN supercell TB Hamiltonian and replace the B and N on-site energies at the substitution sites with the corresponding defect-like unit cell values: from 8.2 eV to 4.37 eV ($sp^2$) and from 4.79 eV to 3.28 eV ($p_z$) for C$_\mathrm{B}$; and from $-1.4$ eV to 0.44 eV ($sp^2$) and from $-2.14$ eV to 0.34 eV ($p_z$) for C$_\mathrm{N}$.

The resulting band structure obtained from the TB supercell (shown in blue in Fig.~\ref{Fig_Case2_BS}) agrees well with the reference DFT supercell band structure (shown in black in Fig.~\ref{Fig_Case2_BS}) in terms of the key qualitative features: the number of in-gap states, their degeneracies, and whether the defect levels are shallow or deep relative to the host band edges. While the absolute energy positions show some quantitative deviations, the essential characteristics relevant for defect screening are reproduced reliably. This agreement is particularly significant because it demonstrates that the unit-cell-based protocol remains effective even when the local point-group symmetry is broken and the hopping variations are substantially larger than in the TMDC cases. As long as the global crystal structure is preserved and the on-site energy shifts continue to play the dominant role in determining the defect state positions, the protocol can capture the key electronic features without requiring explicit supercell DFT calculations for each defect configuration. The C$_\mathrm{B}$C$_\mathrm{N}$ h-BN case thus validates the broader applicability of the approach beyond simple isostructural substitutions.

\subsection{Case study 3: NV\texorpdfstring{$^-$}{-} center in diamond}
\begin{table}[!htb]
    \centering
    \renewcommand{\arraystretch}{1.3}
    \setlength{\tabcolsep}{6pt}
    \begin{tabular}{cc|cc}
    \multirow{2}{*}{Terms} & \multirow{2}{*}{Symmetry}
                        &Diamond        &CN$_\mathrm{C}$  \\
    &                   &               &(N substitutes C)  \\[0.1em]
    \hline
    \multirow{4}{*}{$\varepsilon^{\alpha}$} 
      &\multirow{2}{*}{$s$}
                        &-2.79          &-0.97              \\
      &                 &               &(+1.82)            \\[0.2em]
      &\multirow{2}{*}{$p$}
                        &1.97           &4.37               \\
      &                 &               &(+2.4)             \\[0.2em]
    \multirow{4}{*}{$\varepsilon^{\beta}$} 
      &\multirow{2}{*}{$s$}
                        &-2.79          &-8.57              \\
      &                 &               &(-5.78)            \\[0.2em]
      &\multirow{2}{*}{$p$}
                        &1.97           &1.27               \\
      &                 &               &(-0.7)             \\[0.2em]
    \hline
    \multirow{4}{*}{$t^{\alpha\text{--}\beta}_{i\mu,j\nu}$}
        & \multirow{2}{*}{$s$--$s$}
                        &17.41          &13.15              \\
        &               &               &(-4.26)            \\
        & \multirow{2}{*}{$p$--$p$}
                        &2.79           &2.79               \\
        &               &               &(+0)               \\
    \hline
    \multirow{4}{*}{$t^{\alpha\text{--}\alpha}_{i\mu,j\mu}$}
      &\multirow{2}{*}{$s$--$s$}
                        &0.01           &0.04               \\
      &                 &               &(+0.03)            \\[0.2em]
      &\multirow{2}{*}{$p$--$p$}
                        &0.03           &0.02               \\
      &                 &               &(-0.01)            \\[0.2em]
    \multirow{4}{*}{$t^{\beta\text{--}\beta}_{i\mu,j\mu}$}
      &\multirow{2}{*}{$s$--$s$}
                        &0.01           &0.08                \\
      &                 &               &(+0.07)            \\[0.2em]
      &\multirow{2}{*}{$p$--$p$}
                        &0.03           &0.01               \\
      &                 &               &(-0.02)            \\[0.2em]
    \hline
    \end{tabular}       
    \caption{The first block lists the on-site energies $\varepsilon^\alpha$ and $\varepsilon^\beta$, where $\alpha$ denotes the first atomic site (C in diamond, C in CN$_\mathrm{C}$) and $\beta$ denotes the second atomic site (C in diamond, N in CN$_\mathrm{C}$). Each atomic site is described by one $s$-orbital and three $p$-orbital. All on-site energies are aligned with respect to the valence-band maximum (VBM) of the pristine diamond unit cell. The second block ($t^{\alpha\text{--}\beta}_{i\mu,j\nu}$) presents hopping amplitudes between different orbitals, listed as absolute values. The third block ($t^{\alpha\text{--}\alpha}_{i\mu,j\mu}$ and $t^{\beta\text{--}\beta}_{i\nu,j\nu}$) presents hopping amplitudes between same orbitals at different sites, considering only the first nearest-neighbor shell; these values are averaged to account for direction-dependent variations under symmetry. The values in parentheses indicate the deviation from the corresponding diamond unit-cell Hamiltonian values.}
    \label{Case study 3}
\end{table}

\begin{figure}[t]
    \centering
    \includegraphics[width=0.2\textwidth]{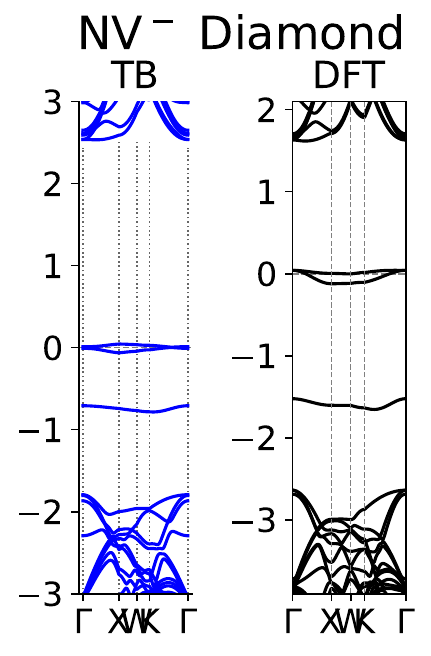}
    \caption{Band structures of NV$^-$ center diamond. The left (blue) bands are obtained from the unit-cell-based protocol, and the right (black) bands are calculated from DFT supercell calculations. The Fermi energy is set at $E_F = 0\ eV$.}
    \label{Fig_Case3_BS}
\end{figure}

The negatively charged nitrogen-vacancy (NV$^-$) center in diamond is one of the most widely studied point defects in solid-state quantum systems. It exhibits a zero-phonon line at 637 nm (1.945 eV), long spin coherence times, and optical addressability of its spin states. These properties have enabled applications in quantum sensing, magnetometry, and quantum information processing. \cite{NV-SPE1, NV-SPE2, NV-SPE3, NV-SPE4, NV-SPE5, NV-SPE6, NV-SPE7}

Structurally, the NV$^-$ center consists of a substitutional nitrogen atom and an adjacent carbon vacancy in the diamond lattice. The pristine diamond lattice has a face-centered cubic structure with tetrahedral $T_\mathrm{d}$ symmetry at each carbon site, where each atom forms four $sp^3$-hybridized covalent bonds with its nearest neighbors. The introduction of the nitrogen substitution and the adjacent vacancy reduces the local symmetry to $C_\mathrm{3v}$, breaking the tetrahedral coordination and leaving three carbon atoms surrounding the vacancy with dangling bonds. This configuration creates a highly localized electronic structure with characteristic defect states deep within the wide band gap of diamond.

The NV$^-$ center serves as the third case study to test the unit-cell-based protocol for several reasons. First, unlike the 2D materials examined previously, diamond is a 3D bulk crystal, requiring a full 3D TB model and introducing additional complexity in constructing the supercell Hamiltonian in real space. Second, the NV$^-$ defect is not a simple substitutional defect but a substitution-vacancy complex, which possibly alters overlaps between orbitals. Third, the local symmetry reduction from $T_\mathrm{d}$ to $C_\mathrm{3v}$ is more severe than in the h-BN case, creating a highly anisotropic bonding environment around the vacancy. These features make the NV$^-$ center an important benchmark for assessing whether the unit-cell-based approach can extend to complex defect configurations in 3D wide-gap semiconductors.

To model the NV$^-$ center, we compute tight-binding Hamiltonians for the host diamond unit cell and a defect-like unit cell CN$_\mathrm{C}$ (where nitrogen substitutes carbon adjacent to a vacancy). The diamond unit cell is constructed as a two-atom primitive cell to minimize computational cost while retaining the essential symmetry and bonding characteristics of the diamond lattice. To maintain a consistent orbital basis across both systems, each atomic site is described by one $s$-orbital and three $p$-orbitals. The TB Hamiltonian parameters extracted from these unit cells are summarized in Table~\ref{Case study 3}. As in the h-BN case, all on-site energies are aligned with respect to the VBM of the pristine diamond unit cell. This alignment strategy is appropriate for wide-gap $sp$-bonded semiconductors, where aligning to the Fermi level would introduce artifacts due to charge pinning near the CBM. As in the previous cases, all hopping amplitudes are presented in absolute values. For same-orbital hopping between different sites ($t^{\alpha\text{--}\alpha}_{i\mu,j\mu}$), only the first nearest-neighbor shell is considered, and the values are averaged to account for direction-dependent variations under symmetry. The values in parentheses indicate the deviation from the corresponding diamond unit cell Hamiltonian values.

From the TB Hamiltonian data in Table~\ref{Case study 3}, the on-site energy shifts at the nitrogen site are $-5.78$ eV ($s$) and $-0.70$ eV ($p$) relative to carbon in diamond. The nearest-neighbor C--N hopping amplitudes also show notable changes: the $s$-$s$ hopping decreases by $-4.26$ eV compared to the C--C bond in pristine diamond, while the $p$-$p$ hopping remains essentially unchanged. Unlike the TMDC and h-BN cases, the $s$-$s$ hopping variation is comparable in magnitude to the on-site shifts rather than an order of magnitude smaller.

This larger relative change in the $s$-$s$ hopping reflects the directional $\sigma$-bonding character of the $sp^3$-hybridized diamond lattice, where the $s$ orbital plays a central role in forming the strong covalent C--C bonds. The introduction of nitrogen, with its higher electronegativity, shifts the bonding from symmetric covalent to partially polar, reducing the $s$-$s$ orbital overlap.

Despite the comparable magnitude of the $s$-$s$ hopping change, we again restrict the modification to the on-site energies when constructing the NV$^-$ supercell Hamiltonian. In the TB formalism, a vacancy is represented by an extremely large on-site energy at the missing site, effectively removing it from the electronic structure. Combined with the nitrogen substitution, the on-site energy shifts of both at the nitrogen and the vacancy site play a dominant role in determining the energy positions of the defect states. Also, as in the previous cases, the protocol's central goal is to capture the key qualitative features of defect electronic structure in a computationally inexpensive manner. In practice, we start from the host diamond supercell TB Hamiltonian and replace the C on-site energies at the substitution site, $-2.79$ eV ($s$) and $1.97$ eV ($p$), with the N on-site energies from the CN$_\mathrm{C}$ defect-like unit cell values, $-8.57$ eV ($s$) and $1.27$ eV ($p$). To construct the vacancy, we remove all Hamiltonian matrix elements associated with one carbon site adjacent to the nitrogen substitution, effectively eliminating that site from the electronic structure.

The resulting band structure obtained from the TB supercell (shown in blue in Fig.~\ref{Fig_Case3_BS}) reproduces the key qualitative features of the NV$^-$ center, although the absolute energy positions of the defect levels show somewhat larger deviations from the reference DFT supercell band structure (shown in black in Fig.~\ref{Fig_Case3_BS}) compared to the TMDC and h-BN cases. This discrepancy arises in part because the unit-cell-based protocol as implemented here captures the electronic structure of the neutral NV$^0$ defect rather than the negatively charged NV$^-$ state. The additional electron in NV$^-$ occupies one of the in-gap defect states and induces substantial electronic relaxation and screening effects that redistribute charge around the vacancy and nitrogen sites, thereby shifting the energies of the defect levels. Because the TB parameters are extracted from charge-neutral unit cells and the protocol does not explicitly account for the self-consistent charge redistribution associated with the extra electron, the predicted defect-level positions correspond more closely to NV$^0$ than to NV$^-$ (See Fig.~\ref{SI_Fig_Case3_BS}).

Despite this limitation, the protocol successfully captures the number of in-gap states, their degeneracies, and their general positions relative to the band edges that is essential for initial defect screening. Most importantly, this case demonstrates that the unit-cell-based approach remains effective even in a challenging scenario where the host material is three-dimensional, the defect involves a vacancy, and the local symmetry is reduced significantly from tetrahedral $T_\mathrm{d}$ to trigonal $C_\mathrm{3v}$. The robustness of the protocol across these three case studies, which are isostructural substitution in TMDCs, symmetry-breaking substitution in h-BN, and substitution-vacancy complexes in 3D diamond, validates its broader applicability for rapid pre-screening of point defects in a wide range of materials systems.

\subsection{Applicability and Limitations of the Protocol}
\begin{figure}[t]
    \centering
    \includegraphics[width=0.4\textwidth]{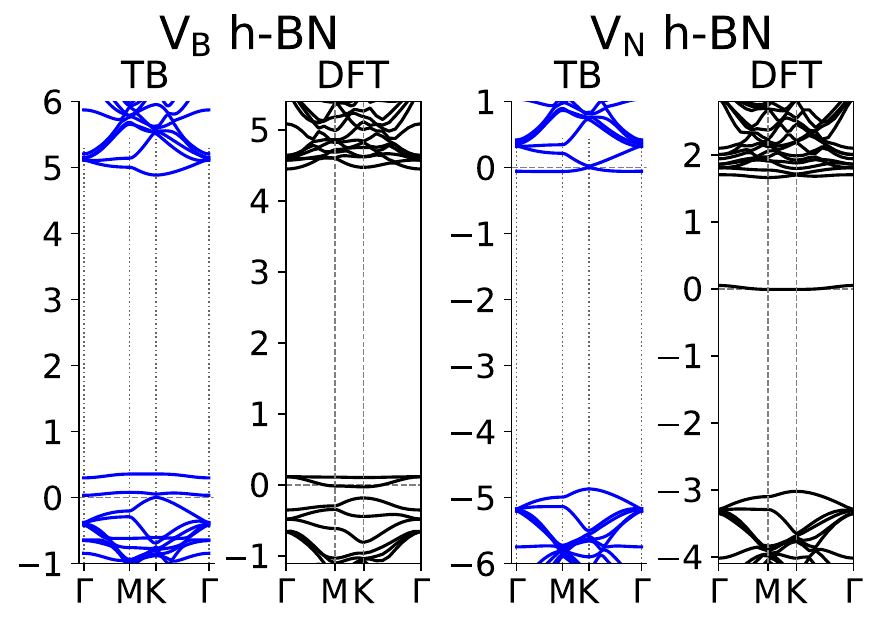}
    \caption{Band structures of V$_\mathrm{B}$ h-BN and V$_\mathrm{N}$ h-BN. The left (blue) bands are obtained from the unit-cell-based protocol, and the right (black) bands are calculated from DFT supercell calculations. The Fermi energy is set at $E_F = 0\ eV$.}
    \label{Limit_Case_BS}
\end{figure}
As demonstrated in the three case studies, the unit-cell-based TB protocol performs robustly for substitutional defects across systems with diverse dimensionalities and local symmetries, provided that the global crystal structure is preserved. In the TMDC case, both global and local point-group symmetry are maintained upon substitution; in h-BN, the global honeycomb lattice is preserved while the local inversion symmetry is broken; and in diamond, the three-dimensional structure is retained despite the introduction of a nitrogen substitution and an adjacent vacancy that reduces the local symmetry from $T_\mathrm{d}$ to $C_\mathrm{3v}$. Across all three systems, modifying only the on-site energies at the defect sites proves sufficient to capture the key qualitative features of the defect electronic structure. This success can be attributed to the fact that, although the hopping parameters change in response to the different atomic species and bonding characteristics, the substitutional defects do not induce large geometric distortions of the host lattice; consequently, the relative magnitude ordering of the hopping elements remains largely unchanged. As a result, the positions and characters of the defect states continue to be dominated by the shifts in on-site energies.

However, the protocol exhibits clear limitations for certain vacancy defects. While the NV$^-$ center in diamond and vacancies of less electronegative atoms (V$_\mathrm{B}$ h-BN) reasonably describe characteristics of defect states well, vacancies of highly electronegative atoms (V$_\mathrm{N}$ h-BN) cannot capture characteristics of defect states well (Fig.~\ref{Limit_Case_BS}). This failure occurs because highly electronegative atoms strongly attract and localize valence electrons from their neighbors, resulting in an electrostatic potential distribution that reflects this charge imbalance. The TB Hamiltonians extracted from relaxed unit cells encode this equilibrium charge distribution. When a highly electronegative atom is removed to create a vacancy, the surrounding atoms lose the strong electron-withdrawing center, and their charge densities must redistribute substantially. This redistribution induces significant shifts in the local electrostatic potential, which in turn modify both the on-site energies and the hopping amplitudes of the neighboring atoms. Because the unit-cell-based protocol simply removes the Hamiltonian matrix elements of the vacancy site without accounting for the self-consistent charge redistribution and screening effects on the surrounding atoms, it fails to capture these potential shifts, leading to inaccurate predictions of the defect state positions. In contrast, vacancies of less electronegative atoms produce more moderate charge redistribution, and removing the Hamiltonian matrix elements of the vacancy site remains a reasonable approximation.

A second important limitation concerns spin polarization and charge-state effects. Throughout this work, the TB parameters have been extracted from non-spin-polarized, charge-neutral unit-cell calculations. As illustrated by the NV$^-$ case, the addition or removal of electrons, or introducing spin polarization, induces electronic relaxation and screening that redistribute charge around the defect site and shift the energies of the defect levels. The unit-cell-based protocol does not account for these self-consistent effects, so the predicted defect-level positions correspond to the neutral, non-spin-polarized configurations. For defects whose properties depend strongly on charge state or spin, the protocol can identify the presence and symmetry of in-gap states but cannot accurately predict their energies in charged or magnetic configurations. Nevertheless, because the primary goal is rapid pre-screening to identify promising defect candidates, this limitation does not undermine the protocol's utility. Candidate defects can be refined through targeted supercell DFT calculations or more sophisticated many-body approaches once identified.

\section{Conclusion}
We have presented a unit-cell-based TB protocol for rapid pre-screening of substitutional point defects in crystalline materials. The workflow proceeds as follows: first, both the host material and defect-like unit cells (with the substituting atom in the same crystal structure as the host) are relaxed using DFT. Second, localized Wannier functions are constructed for each relaxed unit cell, and the corresponding TB Hamiltonians are extracted via Wannier90. Third, the host unit cell Hamiltonian is replicated in real space to construct a supercell TB model of the pristine material. Finally, the defect is introduced by replacing the on-site energies at the substitution site with those from the defect-like unit cell, while leaving the hopping parameters unchanged. This simple modification captures the key qualitative features of defect electronic structures, including the number of in-gap states, their degeneracies, and whether they are shallow or deep relative to the host band edges, without requiring explicit DFT supercell calculations for each defect configuration.

The underlying principle is that substitutional defects in systems where the global crystal structure is preserved induce changes primarily in the on-site terms rather than in the hopping terms. Because the defect-like unit cells are fully relaxed to match the local bonding geometry around the substituting atom, their TB on-site energies already incorporate the chemical and hybridization differences introduced by the defect. By transferring only these on-site parameters to the supercell model, the protocol efficiently approximates the defect electronic structure at a fraction of the computational cost.

The protocol has been validated across three diverse case studies: isostructural substitutional defects in 2D TMDC (M$_\mathrm{Mo}$ MoS$_2$, M = Ce, Zr, Nb, Tc, and Ru), symmetry-breaking carbon substitutions in hexagonal boron nitride (C$_\mathrm{B}$C$_\mathrm{N}$ h-BN), and the nitrogen-vacancy (NV$^-$) center in 3D diamond. These systems test the protocol's robustness under varying conditions: 2D versus 3D hosts, preservation versus breaking of local point-group symmetry, and simple substitution versus substitution-vacancy complexes. In all cases, the TB supercell band structures agree well with reference DFT supercell results in terms of the essential characteristics relevant for defect screening, despite some quantitative deviations in the absolute energy positions of defect levels.

However, the protocol also exhibits important limitations. Vacancies of highly electronegative atoms induce substantial charge redistribution and screening effects on surrounding atoms that cannot be captured by simply removing the vacancy site Hamiltonian elements. Similarly, the protocol does not account for self-consistent electronic relaxation and spin polarization associated with charged or magnetic defect configurations, limiting its quantitative accuracy for such systems. 

Nevertheless, for the primary purpose of rapidly identifying promising defect candidates from a large pool of possibilities, the protocol provides a practical and scalable tool for high-throughput defect screening. By greatly reducing the computational cost of initial defect exploration, it enables systematic surveys of large defect spaces and enables the discovery of new point defects with new functionalities.

\section*{Data Availability}

\bibliography{ref}

\providecommand{\latin}[1]{#1}
\makeatletter
\providecommand{\doi}
  {\begingroup\let\do\@makeother\dospecials
  \catcode`\{=1 \catcode`\}=2 \doi@aux}
\providecommand{\doi@aux}[1]{\endgroup\texttt{#1}}
\makeatother
\providecommand*\mcitethebibliography{\thebibliography}
\csname @ifundefined\endcsname{endmcitethebibliography}  {\let\endmcitethebibliography\endthebibliography}{}
\begin{mcitethebibliography}{65}
\providecommand*\natexlab[1]{#1}
\providecommand*\mciteSetBstSublistMode[1]{}
\providecommand*\mciteSetBstMaxWidthForm[2]{}
\providecommand*\mciteBstWouldAddEndPuncttrue
  {\def\EndOfBibitem{\unskip.}}
\providecommand*\mciteBstWouldAddEndPunctfalse
  {\let\EndOfBibitem\relax}
\providecommand*\mciteSetBstMidEndSepPunct[3]{}
\providecommand*\mciteSetBstSublistLabelBeginEnd[3]{}
\providecommand*\EndOfBibitem{}
\mciteSetBstSublistMode{f}
\mciteSetBstMaxWidthForm{subitem}{(\alph{mcitesubitemcount})}
\mciteSetBstSublistLabelBeginEnd
  {\mcitemaxwidthsubitemform\space}
  {\relax}
  {\relax}

\bibitem[Mahajan(2000)]{point_defect_effects_1}
Mahajan,~S. Defects in semiconductors and their effects on devices. \emph{Acta Mater.} \textbf{2000}, \emph{48}, 137--149\relax
\mciteBstWouldAddEndPuncttrue
\mciteSetBstMidEndSepPunct{\mcitedefaultmidpunct}
{\mcitedefaultendpunct}{\mcitedefaultseppunct}\relax
\EndOfBibitem
\bibitem[Dreyer \latin{et~al.}(2024)Dreyer, Janotti, Lyons, and Wickramaratne]{point_defect_effects_2}
Dreyer,~C.~E.; Janotti,~A.; Lyons,~J.~L.; Wickramaratne,~D. Defects in semiconductors. \emph{J. Appl. Phys.} \textbf{2024}, \emph{136}, 190401\relax
\mciteBstWouldAddEndPuncttrue
\mciteSetBstMidEndSepPunct{\mcitedefaultmidpunct}
{\mcitedefaultendpunct}{\mcitedefaultseppunct}\relax
\EndOfBibitem
\bibitem[Koehl \latin{et~al.}(2015)Koehl, Seo, Galli, and Awschalom]{point_defect_effects_3}
Koehl,~W.~F.; Seo,~H.; Galli,~G.; Awschalom,~D.~D. Designing Defect Spins for Wafer-Scale Quantum Technologies. \emph{MRS Bull.} \textbf{2015}, \emph{40}, 1146--1153\relax
\mciteBstWouldAddEndPuncttrue
\mciteSetBstMidEndSepPunct{\mcitedefaultmidpunct}
{\mcitedefaultendpunct}{\mcitedefaultseppunct}\relax
\EndOfBibitem
\bibitem[Loredo \latin{et~al.}(2016)Loredo, Zakaria, Somaschi, Anton, de~Santis, Giesz, Grange, Broome, Gazzano, Coppola, Sagnes, Lemaitre, Auffeves, Senellart, Almeida, and White]{point_defect_effects_4}
Loredo,~J.~C.; Zakaria,~N.~A.; Somaschi,~N.; Anton,~C.; de~Santis,~L.; Giesz,~V.; Grange,~T.; Broome,~M.~A.; Gazzano,~O.; Coppola,~G. \latin{et~al.}  Scalable Performance in Solid-State Single-Photon Sources. \emph{Optica} \textbf{2016}, \emph{3}, 433--440\relax
\mciteBstWouldAddEndPuncttrue
\mciteSetBstMidEndSepPunct{\mcitedefaultmidpunct}
{\mcitedefaultendpunct}{\mcitedefaultseppunct}\relax
\EndOfBibitem
\bibitem[Somaschi \latin{et~al.}(2016)Somaschi, Giesz, De~Santis, Loredo, Almeida, Hornecker, Portalupi, Grange, Antón, Demory, Gomez~Carbonell, Sagnes, Lanzillotti~Kimura, Lemaître, Auffèves, White, Lanco, and Senellart]{point_defect_effects_5}
Somaschi,~N.; Giesz,~V.; De~Santis,~L.; Loredo,~J.; Almeida,~M.; Hornecker,~G.; Portalupi,~S.; Grange,~T.; Antón,~C.; Demory,~J. \latin{et~al.}  Near-Optimal Single-Photon Sources in the Solid State. \emph{Nat. Photon.} \textbf{2016}, \emph{10}, 340--345\relax
\mciteBstWouldAddEndPuncttrue
\mciteSetBstMidEndSepPunct{\mcitedefaultmidpunct}
{\mcitedefaultendpunct}{\mcitedefaultseppunct}\relax
\EndOfBibitem
\bibitem[Ding \latin{et~al.}(2016)Ding, He, Duan, Gregersen, Chen, Unsleber, Maier, Schneider, Kamp, H\"ofling, Lu, and Pan]{point_defect_effects_6}
Ding,~X.; He,~Y.; Duan,~Z.-C.; Gregersen,~N.; Chen,~M.-C.; Unsleber,~S.; Maier,~S.; Schneider,~C.; Kamp,~M.; H\"ofling,~S. \latin{et~al.}  On-Demand Single Photons with High Extraction Efficiency and Near-Unity Indistinguishability from a Resonantly Driven Quantum Dot in a Micropillar. \emph{Phys. Rev. Lett.} \textbf{2016}, \emph{116}, 020401\relax
\mciteBstWouldAddEndPuncttrue
\mciteSetBstMidEndSepPunct{\mcitedefaultmidpunct}
{\mcitedefaultendpunct}{\mcitedefaultseppunct}\relax
\EndOfBibitem
\bibitem[Kane(1998)]{PDforquantum1}
Kane,~B.~E. A Silicon-Based Nuclear Spin Quantum Computer. \emph{Nature} \textbf{1998}, \emph{393}, 133--137\relax
\mciteBstWouldAddEndPuncttrue
\mciteSetBstMidEndSepPunct{\mcitedefaultmidpunct}
{\mcitedefaultendpunct}{\mcitedefaultseppunct}\relax
\EndOfBibitem
\bibitem[Weber \latin{et~al.}(2010)Weber, Koehl, Varley, Janotti, Buckley, Walle, and Awschalom]{PDforquantum2}
Weber,~J.; Koehl,~W.; Varley,~J.; Janotti,~A.; Buckley,~B.; Walle,~C.; Awschalom,~D. Quantum Computing with Defects. \emph{PNAS} \textbf{2010}, \emph{107}, 8513--8518\relax
\mciteBstWouldAddEndPuncttrue
\mciteSetBstMidEndSepPunct{\mcitedefaultmidpunct}
{\mcitedefaultendpunct}{\mcitedefaultseppunct}\relax
\EndOfBibitem
\bibitem[Aharonovich \latin{et~al.}(2011)Aharonovich, Castelletto, Simpson, Su, Greentree, and Prawer]{PDforquantum3}
Aharonovich,~I.; Castelletto,~S.; Simpson,~D.~A.; Su,~C.-H.; Greentree,~A.~D.; Prawer,~S. Diamond-Based Single-Photon Emitters. \emph{Rep. Prog. Phys.} \textbf{2011}, \emph{74}, 076501\relax
\mciteBstWouldAddEndPuncttrue
\mciteSetBstMidEndSepPunct{\mcitedefaultmidpunct}
{\mcitedefaultendpunct}{\mcitedefaultseppunct}\relax
\EndOfBibitem
\bibitem[Pla \latin{et~al.}(2012)Pla, Tan, Dehollain, Lim, Morton, Jamieson, Dzurak, and Morello]{PDforquantum4}
Pla,~J.; Tan,~K.; Dehollain,~J.~P.; Lim,~W.; Morton,~J.; Jamieson,~D.; Dzurak,~A.; Morello,~A. A single-Atom Electron Spin Qubit in Silicon. \emph{Nature} \textbf{2012}, \emph{489}, 541--545\relax
\mciteBstWouldAddEndPuncttrue
\mciteSetBstMidEndSepPunct{\mcitedefaultmidpunct}
{\mcitedefaultendpunct}{\mcitedefaultseppunct}\relax
\EndOfBibitem
\bibitem[Wu \latin{et~al.}(2019)Wu, Wang, Qin, Rong, and Du]{PDforquantum5}
Wu,~Y.; Wang,~Y.; Qin,~X.; Rong,~X.; Du,~J. A programmable Two-Qubit Solid-State Quantum Processor under Ambient Conditions. \emph{Npj Quantum Inf.} \textbf{2019}, \emph{5}, 9\relax
\mciteBstWouldAddEndPuncttrue
\mciteSetBstMidEndSepPunct{\mcitedefaultmidpunct}
{\mcitedefaultendpunct}{\mcitedefaultseppunct}\relax
\EndOfBibitem
\bibitem[Muechler \latin{et~al.}(2020)Muechler, Hu, Lin, Yang, and Car]{PDforquantum6}
Muechler,~L.; Hu,~W.; Lin,~L.; Yang,~C.; Car,~R. Influence of Point Defects on the Electronic and Topological Properties of Monolayer $\mathrm{{WTe}}_{\mathrm{{2}}}$. \emph{Phys. Rev. B} \textbf{2020}, \emph{102}, 041103\relax
\mciteBstWouldAddEndPuncttrue
\mciteSetBstMidEndSepPunct{\mcitedefaultmidpunct}
{\mcitedefaultendpunct}{\mcitedefaultseppunct}\relax
\EndOfBibitem
\bibitem[Aharonovich and Neu(2014)Aharonovich, and Neu]{NV-SPE1}
Aharonovich,~I.; Neu,~E. Diamond Nanophotonics. \emph{Adv. Opt. Mater.} \textbf{2014}, \emph{2}, 911--928\relax
\mciteBstWouldAddEndPuncttrue
\mciteSetBstMidEndSepPunct{\mcitedefaultmidpunct}
{\mcitedefaultendpunct}{\mcitedefaultseppunct}\relax
\EndOfBibitem
\bibitem[Alkauskas \latin{et~al.}(2014)Alkauskas, Buckley, Awschalom, and de~Walle]{NV-SPE2}
Alkauskas,~A.; Buckley,~B.~B.; Awschalom,~D.~D.; de~Walle,~C. G.~V. First-Principles Theory of the Luminescence Lineshape for the Triplet Transition in Diamond NV Centres. \emph{New J. Phys.} \textbf{2014}, \emph{16}, 073026\relax
\mciteBstWouldAddEndPuncttrue
\mciteSetBstMidEndSepPunct{\mcitedefaultmidpunct}
{\mcitedefaultendpunct}{\mcitedefaultseppunct}\relax
\EndOfBibitem
\bibitem[Jelezko \latin{et~al.}(2004)Jelezko, Gaebel, Popa, Gruber, and Wrachtrup]{NV-SPE3}
Jelezko,~F.; Gaebel,~T.; Popa,~I.; Gruber,~A.; Wrachtrup,~J. Observation of Coherent Oscillations in a Single Electron Spin. \emph{Phys. Rev. Lett.} \textbf{2004}, \emph{92}, 076401\relax
\mciteBstWouldAddEndPuncttrue
\mciteSetBstMidEndSepPunct{\mcitedefaultmidpunct}
{\mcitedefaultendpunct}{\mcitedefaultseppunct}\relax
\EndOfBibitem
\bibitem[Jelezko \latin{et~al.}(2004)Jelezko, Gaebel, Popa, Domhan, Gruber, and Wrachtrup]{NV-SPE4}
Jelezko,~F.; Gaebel,~T.; Popa,~I.; Domhan,~M.; Gruber,~A.; Wrachtrup,~J. Observation of Coherent Oscillation of a Single Nuclear Spin and Realization of a Two-Qubit Conditional Quantum Gate. \emph{Phys. Rev. Lett.} \textbf{2004}, \emph{93}, 130501\relax
\mciteBstWouldAddEndPuncttrue
\mciteSetBstMidEndSepPunct{\mcitedefaultmidpunct}
{\mcitedefaultendpunct}{\mcitedefaultseppunct}\relax
\EndOfBibitem
\bibitem[Balasubramanian \latin{et~al.}(2009)Balasubramanian, Neumann, Twitchen, Markham, Kolesov, Mizuochi, Isoya, Achard, Beck, Tissler, Jacques, Hemmer, and Jelezko]{NV-SPE5}
Balasubramanian,~G.; Neumann,~P.; Twitchen,~D.; Markham,~M.; Kolesov,~R.; Mizuochi,~N.; Isoya,~J.; Achard,~J.; Beck,~J.; Tissler,~J. \latin{et~al.}  Ultralong spin coherence time in isotopically engineered diamond. \emph{Nat. Mater.} \textbf{2009}, \emph{8}, 383--387\relax
\mciteBstWouldAddEndPuncttrue
\mciteSetBstMidEndSepPunct{\mcitedefaultmidpunct}
{\mcitedefaultendpunct}{\mcitedefaultseppunct}\relax
\EndOfBibitem
\bibitem[Doherty \latin{et~al.}(2011)Doherty, Manson, Delaney, and Hollenberg]{NV-SPE6}
Doherty,~M.~W.; Manson,~N.~B.; Delaney,~P.; Hollenberg,~L. C.~L. The negatively charged nitrogen-vacancy centre in diamond: the electronic solution. \emph{New J. Phys.} \textbf{2011}, \emph{13}, 025019\relax
\mciteBstWouldAddEndPuncttrue
\mciteSetBstMidEndSepPunct{\mcitedefaultmidpunct}
{\mcitedefaultendpunct}{\mcitedefaultseppunct}\relax
\EndOfBibitem
\bibitem[Maze \latin{et~al.}(2011)Maze, Gali, Togan, Chu, Trifonov, Kaxiras, and Lukin]{NV-SPE7}
Maze,~J.~R.; Gali,~A.; Togan,~E.; Chu,~Y.; Trifonov,~A.; Kaxiras,~E.; Lukin,~M.~D. Properties of nitrogen-vacancy centers in diamond: the group theoretic approach. \emph{New J. Phys.} \textbf{2011}, \emph{13}, 025025\relax
\mciteBstWouldAddEndPuncttrue
\mciteSetBstMidEndSepPunct{\mcitedefaultmidpunct}
{\mcitedefaultendpunct}{\mcitedefaultseppunct}\relax
\EndOfBibitem
\bibitem[Tang \latin{et~al.}(2025)Tang, Ritika, Tamtaji, Liu, Hu, Liu, Galligan, Xu, Shen, Wang, You, Li, Chen, Aharonovich, and Luo]{CBCN_h-BN1}
Tang,~T.~W.; Ritika,~R.; Tamtaji,~M.; Liu,~H.; Hu,~Y.; Liu,~Z.; Galligan,~P.~R.; Xu,~M.; Shen,~J.; Wang,~J. \latin{et~al.}  Structured-Defect Engineering of Hexagonal Boron Nitride for Identified Visible Single-Photon Emitters. \emph{ACS Nano} \textbf{2025}, \emph{19}, 8509--8519\relax
\mciteBstWouldAddEndPuncttrue
\mciteSetBstMidEndSepPunct{\mcitedefaultmidpunct}
{\mcitedefaultendpunct}{\mcitedefaultseppunct}\relax
\EndOfBibitem
\bibitem[Cholsuk \latin{et~al.}(2024)Cholsuk, Zand, {\c{C}}akan, and Vogl]{CBCN_h-BN2}
Cholsuk,~C.; Zand,~A.; {\c{C}}akan,~A.; Vogl,~T. The hBN Defects Database: A Theoretical Compilation of Color Centers in Hexagonal Boron Nitride. \emph{J. Phys. Chem. C} \textbf{2024}, \emph{128}, 12716--12725\relax
\mciteBstWouldAddEndPuncttrue
\mciteSetBstMidEndSepPunct{\mcitedefaultmidpunct}
{\mcitedefaultendpunct}{\mcitedefaultseppunct}\relax
\EndOfBibitem
\bibitem[Du \latin{et~al.}(2025)Du, Yang, Pei, Han, Feng, Zhao, and Zhou]{CBCN_h-BN3}
Du,~T.; Yang,~X.; Pei,~W.; Han,~P.; Feng,~Y.; Zhao,~J.; Zhou,~S. Carbon defects engineering in hexagonal boron nitride for single photon emission and spin qubits. \emph{APL Quantum} \textbf{2025}, \emph{2}, 026127\relax
\mciteBstWouldAddEndPuncttrue
\mciteSetBstMidEndSepPunct{\mcitedefaultmidpunct}
{\mcitedefaultendpunct}{\mcitedefaultseppunct}\relax
\EndOfBibitem
\bibitem[Era \latin{et~al.}(1981)Era, Minami, and Kuzuba]{CBCN_h-BN4}
Era,~K.; Minami,~F.; Kuzuba,~T. Fast luminescence from carbon-related defects of hexagonal boron nitride. \emph{J. Lumin.} \textbf{1981}, \emph{24-25}, 71--74\relax
\mciteBstWouldAddEndPuncttrue
\mciteSetBstMidEndSepPunct{\mcitedefaultmidpunct}
{\mcitedefaultendpunct}{\mcitedefaultseppunct}\relax
\EndOfBibitem
\bibitem[Du \latin{et~al.}(2015)Du, Li, Lin, and Jiang]{CBCN_h-BN5}
Du,~X.~Z.; Li,~J.; Lin,~J.~Y.; Jiang,~H.~X. The origin of deep-level impurity transitions in hexagonal boron nitride. \emph{Appl. Phys. Lett.} \textbf{2015}, \emph{106}, 021110\relax
\mciteBstWouldAddEndPuncttrue
\mciteSetBstMidEndSepPunct{\mcitedefaultmidpunct}
{\mcitedefaultendpunct}{\mcitedefaultseppunct}\relax
\EndOfBibitem
\bibitem[Mackoit-Sinkevičienė \latin{et~al.}(2019)Mackoit-Sinkevičienė, Maciaszek, Van~de Walle, and Alkauskas]{CBCN_h-BN6}
Mackoit-Sinkevičienė,~M.; Maciaszek,~M.; Van~de Walle,~C.~G.; Alkauskas,~A. Carbon dimer defect as a source of the 4.1 eV luminescence in hexagonal boron nitride. \emph{Appl. Phys. Lett.} \textbf{2019}, \emph{115}, 212101\relax
\mciteBstWouldAddEndPuncttrue
\mciteSetBstMidEndSepPunct{\mcitedefaultmidpunct}
{\mcitedefaultendpunct}{\mcitedefaultseppunct}\relax
\EndOfBibitem
\bibitem[Jara \latin{et~al.}(2021)Jara, Rauch, Botti, Marques, Norambuena, Coto, Castellanos-Águila, Maze, and Munoz]{CBCN_h-BN7}
Jara,~C.; Rauch,~T.; Botti,~S.; Marques,~M. A.~L.; Norambuena,~A.; Coto,~R.; Castellanos-Águila,~J.~E.; Maze,~J.~R.; Munoz,~F. First-Principles Identification of Single Photon Emitters Based on Carbon Clusters in Hexagonal Boron Nitride. \emph{J. Phys. Chem. A} \textbf{2021}, \emph{125}, 1325--1335\relax
\mciteBstWouldAddEndPuncttrue
\mciteSetBstMidEndSepPunct{\mcitedefaultmidpunct}
{\mcitedefaultendpunct}{\mcitedefaultseppunct}\relax
\EndOfBibitem
\bibitem[Schell \latin{et~al.}(2017)Schell, Takashima, Tran, Aharonovich, and Takeuchi]{TMDCs1}
Schell,~A.; Takashima,~H.; Tran,~T.; Aharonovich,~I.; Takeuchi,~S. Coupling Quantum Emitters in 2D Materials with Tapered Fibers. \emph{ACS Photonics} \textbf{2017}, \emph{4}, 761--767\relax
\mciteBstWouldAddEndPuncttrue
\mciteSetBstMidEndSepPunct{\mcitedefaultmidpunct}
{\mcitedefaultendpunct}{\mcitedefaultseppunct}\relax
\EndOfBibitem
\bibitem[Tonndorf \latin{et~al.}(2015)Tonndorf, Schmidt, Schneider, Kern, Buscema, Steele, Castellanos-Gomez, van~der Zant, de~Vasconcellos, and Bratschitsch]{TMDCs2}
Tonndorf,~P.; Schmidt,~R.; Schneider,~R.; Kern,~J.; Buscema,~M.; Steele,~G.~A.; Castellanos-Gomez,~A.; van~der Zant,~H. S.~J.; de~Vasconcellos,~S.~M.; Bratschitsch,~R. Single-photon emission from localized excitons in an atomically thin semiconductor. \emph{Optica} \textbf{2015}, \emph{2}, 347--352\relax
\mciteBstWouldAddEndPuncttrue
\mciteSetBstMidEndSepPunct{\mcitedefaultmidpunct}
{\mcitedefaultendpunct}{\mcitedefaultseppunct}\relax
\EndOfBibitem
\bibitem[Wu \latin{et~al.}(2019)Wu, Dass, Hendrickson, Montaño, Fischer, Zhang, Choudhury, Redwing, Wang, and Pettes]{TMDCs3}
Wu,~W.; Dass,~C.~K.; Hendrickson,~J.~R.; Montaño,~R.~D.; Fischer,~R.~E.; Zhang,~X.; Choudhury,~T.~H.; Redwing,~J.~M.; Wang,~Y.; Pettes,~M.~T. Locally defined quantum emission from epitaxial few-layer tungsten diselenide. \emph{Appl. Phys. Lett.} \textbf{2019}, \emph{114}, 213102\relax
\mciteBstWouldAddEndPuncttrue
\mciteSetBstMidEndSepPunct{\mcitedefaultmidpunct}
{\mcitedefaultendpunct}{\mcitedefaultseppunct}\relax
\EndOfBibitem
\bibitem[He \latin{et~al.}(2014)He, Clark, Schaibley, He, Chen, Wei, Ding, Zhang, Yao, Xu, Lu, and Pan]{TMDCs4}
He,~Y.-M.; Clark,~G.; Schaibley,~J.; He,~Y.; Chen,~M.-c.; Wei,~Y.-J.; Ding,~X.; Zhang,~Q.; Yao,~W.; Xu,~X. \latin{et~al.}  Single Quantum Emitters in Monolayer Semiconductors. \emph{Nat. Nanotechnol.} \textbf{2014}, \emph{10}, 497--502\relax
\mciteBstWouldAddEndPuncttrue
\mciteSetBstMidEndSepPunct{\mcitedefaultmidpunct}
{\mcitedefaultendpunct}{\mcitedefaultseppunct}\relax
\EndOfBibitem
\bibitem[Ouma \latin{et~al.}(2017)Ouma, Singh, Obodo, Amolo, and Romero]{LnTMDCs1}
Ouma,~C. N.~M.; Singh,~S.; Obodo,~K.~O.; Amolo,~G.~O.; Romero,~A.~H. Controlling the magnetic and optical responses of a $\mathrm{{MoS}}_{\mathrm{{2}}}$ monolayer by lanthanide substitutional doping: a first-principles study. \emph{Phys. Chem. Chem. Phys.} \textbf{2017}, \emph{19}, 25555--25563\relax
\mciteBstWouldAddEndPuncttrue
\mciteSetBstMidEndSepPunct{\mcitedefaultmidpunct}
{\mcitedefaultendpunct}{\mcitedefaultseppunct}\relax
\EndOfBibitem
\bibitem[López-Morales \latin{et~al.}(2022)López-Morales, Hampel, López, Menon, Flick, and Meriles]{LnTMDCs2}
López-Morales,~G.; Hampel,~A.; López,~G.; Menon,~V.; Flick,~J.; Meriles,~C. Ab-initio investigation of $\mathrm{{Er}}^{\mathrm{{3+}}}$ defects in tungsten disulfide. \emph{Comput. Mater. Sci.} \textbf{2022}, \emph{210}, 111041\relax
\mciteBstWouldAddEndPuncttrue
\mciteSetBstMidEndSepPunct{\mcitedefaultmidpunct}
{\mcitedefaultendpunct}{\mcitedefaultseppunct}\relax
\EndOfBibitem
\bibitem[Maleki-Ghaleh \latin{et~al.}(2024)Maleki-Ghaleh, Moradpur-Tari, Shakiba, Paczesny, Hurley, Siadati, Ansari, and Gity]{LnTMDCs3}
Maleki-Ghaleh,~H.; Moradpur-Tari,~E.; Shakiba,~M.; Paczesny,~J.; Hurley,~P.~K.; Siadati,~M.~H.; Ansari,~L.; Gity,~F. Electronic structure of rare-earth erbium-doped platinum diselenide: A density functional theory study. \emph{J. Phys. Chem. Solids} \textbf{2024}, \emph{190}, 112004\relax
\mciteBstWouldAddEndPuncttrue
\mciteSetBstMidEndSepPunct{\mcitedefaultmidpunct}
{\mcitedefaultendpunct}{\mcitedefaultseppunct}\relax
\EndOfBibitem
\bibitem[Kang \latin{et~al.}(2025)Kang, Queiroz, and Muechler]{LnTMDCs4}
Kang,~H.; Queiroz,~R.; Muechler,~L. Electronic and optical properties of lanthanide-doped ${\mathrm{MoS}}_{2}$: Impact of ionic size and orbital configuration mismatch. \emph{Phys. Rev. B} \textbf{2025}, \emph{112}, 174107\relax
\mciteBstWouldAddEndPuncttrue
\mciteSetBstMidEndSepPunct{\mcitedefaultmidpunct}
{\mcitedefaultendpunct}{\mcitedefaultseppunct}\relax
\EndOfBibitem
\bibitem[Dreyer \latin{et~al.}(2018)Dreyer, Alkauskas, Lyons, Janotti, and Van~de Walle]{SupercellApproach1}
Dreyer,~C.~E.; Alkauskas,~A.; Lyons,~J.~L.; Janotti,~A.; Van~de Walle,~C.~G. First-Principles Calculations of Point Defects for Quantum Technologies. \emph{Annu. Rev. Mater. Res.} \textbf{2018}, \emph{48}, 1--26\relax
\mciteBstWouldAddEndPuncttrue
\mciteSetBstMidEndSepPunct{\mcitedefaultmidpunct}
{\mcitedefaultendpunct}{\mcitedefaultseppunct}\relax
\EndOfBibitem
\bibitem[Freysoldt \latin{et~al.}(2014)Freysoldt, Grabowski, Hickel, Neugebauer, Kresse, Janotti, and Van~de Walle]{SupercellApproach2}
Freysoldt,~C.; Grabowski,~B.; Hickel,~T.; Neugebauer,~J.; Kresse,~G.; Janotti,~A.; Van~de Walle,~C.~G. First-principles calculations for point defects in solids. \emph{Rev. Mod. Phys.} \textbf{2014}, \emph{86}, 253--305\relax
\mciteBstWouldAddEndPuncttrue
\mciteSetBstMidEndSepPunct{\mcitedefaultmidpunct}
{\mcitedefaultendpunct}{\mcitedefaultseppunct}\relax
\EndOfBibitem
\bibitem[Van~de Walle and Neugebauer(2004)Van~de Walle, and Neugebauer]{SupercellApproach3}
Van~de Walle,~C.~G.; Neugebauer,~J. First-principles calculations for defects and impurities: Applications to III-nitrides. \emph{J. Appl. Phys.} \textbf{2004}, \emph{95}, 3851--3879\relax
\mciteBstWouldAddEndPuncttrue
\mciteSetBstMidEndSepPunct{\mcitedefaultmidpunct}
{\mcitedefaultendpunct}{\mcitedefaultseppunct}\relax
\EndOfBibitem
\bibitem[Chen \latin{et~al.}(2025)Chen, Yu, Jin, Govoni, and Galli]{SupercellApproachQE1}
Chen,~S.; Yu,~V. W.-z.; Jin,~Y.; Govoni,~M.; Galli,~G. Advances in Quantum Defect Embedding Theory. \emph{J. Chem. Theory Comput.} \textbf{2025}, \emph{21}, 7797--7812\relax
\mciteBstWouldAddEndPuncttrue
\mciteSetBstMidEndSepPunct{\mcitedefaultmidpunct}
{\mcitedefaultendpunct}{\mcitedefaultseppunct}\relax
\EndOfBibitem
\bibitem[Sun and Chan(2016)Sun, and Chan]{SupercellApproachQE2}
Sun,~Q.; Chan,~G. K.-L. Quantum Embedding Theories. \emph{Acc. Chem. Res.} \textbf{2016}, \emph{49}, 2705--2712\relax
\mciteBstWouldAddEndPuncttrue
\mciteSetBstMidEndSepPunct{\mcitedefaultmidpunct}
{\mcitedefaultendpunct}{\mcitedefaultseppunct}\relax
\EndOfBibitem
\bibitem[Sheng \latin{et~al.}(2022)Sheng, Vorwerk, Govoni, and Galli]{SupercellApproachQE3}
Sheng,~N.; Vorwerk,~C.; Govoni,~M.; Galli,~G. Green’s Function Formulation of Quantum Defect Embedding Theory. \emph{J. Chem. Theory Comput.} \textbf{2022}, \emph{18}, 3512--3522\relax
\mciteBstWouldAddEndPuncttrue
\mciteSetBstMidEndSepPunct{\mcitedefaultmidpunct}
{\mcitedefaultendpunct}{\mcitedefaultseppunct}\relax
\EndOfBibitem
\bibitem[Muechler \latin{et~al.}(2022)Muechler, Badrtdinov, Hampel, Cano, R\"osner, and Dreyer]{SupercellApproachQE4}
Muechler,~L.; Badrtdinov,~D.~I.; Hampel,~A.; Cano,~J.; R\"osner,~M.; Dreyer,~C.~E. Quantum embedding methods for correlated excited states of point defects: Case studies and challenges. \emph{Phys. Rev. B} \textbf{2022}, \emph{105}, 235104\relax
\mciteBstWouldAddEndPuncttrue
\mciteSetBstMidEndSepPunct{\mcitedefaultmidpunct}
{\mcitedefaultendpunct}{\mcitedefaultseppunct}\relax
\EndOfBibitem
\bibitem[Urban \latin{et~al.}(2011)Urban, Reese, Mrovec, Els\"asser, and Meyer]{TB1}
Urban,~A.; Reese,~M.; Mrovec,~M.; Els\"asser,~C.; Meyer,~B. Parameterization of tight-binding models from density functional theory calculations. \emph{Phys. Rev. B} \textbf{2011}, \emph{84}, 155119\relax
\mciteBstWouldAddEndPuncttrue
\mciteSetBstMidEndSepPunct{\mcitedefaultmidpunct}
{\mcitedefaultendpunct}{\mcitedefaultseppunct}\relax
\EndOfBibitem
\bibitem[Singh \latin{et~al.}(1982)Singh, Lindefelt, and Zunger]{TB2}
Singh,~V.~A.; Lindefelt,~U.; Zunger,~A. Evaluation of tight-binding models for deep defect levels in semiconductors. \emph{Phys. Rev. B} \textbf{1982}, \emph{25}, 2781--2785\relax
\mciteBstWouldAddEndPuncttrue
\mciteSetBstMidEndSepPunct{\mcitedefaultmidpunct}
{\mcitedefaultendpunct}{\mcitedefaultseppunct}\relax
\EndOfBibitem
\bibitem[Martin(2012)]{TB3}
Martin,~R.~M. \emph{Electronic Structure: Basic Theory and Practical Methods}; Cambridge University Press: Cambridge, 2012\relax
\mciteBstWouldAddEndPuncttrue
\mciteSetBstMidEndSepPunct{\mcitedefaultmidpunct}
{\mcitedefaultendpunct}{\mcitedefaultseppunct}\relax
\EndOfBibitem
\bibitem[Seifert \latin{et~al.}(1996)Seifert, Porezag, and Frauenheim]{TB4}
Seifert,~G.; Porezag,~D.; Frauenheim,~T. Calculations of molecules, clusters, and solids with a simplified LCAO-DFT-LDA scheme. \emph{Int. J. Quantum Chem.} \textbf{1996}, \emph{58}, 185--192\relax
\mciteBstWouldAddEndPuncttrue
\mciteSetBstMidEndSepPunct{\mcitedefaultmidpunct}
{\mcitedefaultendpunct}{\mcitedefaultseppunct}\relax
\EndOfBibitem
\bibitem[Fried \latin{et~al.}(2025)Fried, Barragan-Yani, Libisch, and Wirtz]{MLApproach1}
Fried,~H.; Barragan-Yani,~D.; Libisch,~F.; Wirtz,~L. A machine learning approach to predict tight-binding parameters for point defects via the projected density of states. \emph{npj Comput. Mater.} \textbf{2025}, \emph{11}\relax
\mciteBstWouldAddEndPuncttrue
\mciteSetBstMidEndSepPunct{\mcitedefaultmidpunct}
{\mcitedefaultendpunct}{\mcitedefaultseppunct}\relax
\EndOfBibitem
\bibitem[Wang \latin{et~al.}(2021)Wang, Shizhuo, Wang, He, Huang, and Chang]{MLApproach2}
Wang,~Z.; Shizhuo,~Y.; Wang,~H.; He,~J.; Huang,~Q.; Chang,~S. Machine learning method for tight-binding Hamiltonian parameterization from ab-initio band structure. \emph{npj Comput. Mater.} \textbf{2021}, \emph{7}\relax
\mciteBstWouldAddEndPuncttrue
\mciteSetBstMidEndSepPunct{\mcitedefaultmidpunct}
{\mcitedefaultendpunct}{\mcitedefaultseppunct}\relax
\EndOfBibitem
\bibitem[Nakhaee \latin{et~al.}(2020)Nakhaee, Ketabi, and Peeters]{MLApproach3}
Nakhaee,~M.; Ketabi,~S.~A.; Peeters,~F.~M. Machine learning approach to constructing tight binding models for solids with application to BiTeCl. \emph{J. Appl. Phys.} \textbf{2020}, \emph{128}, 215107\relax
\mciteBstWouldAddEndPuncttrue
\mciteSetBstMidEndSepPunct{\mcitedefaultmidpunct}
{\mcitedefaultendpunct}{\mcitedefaultseppunct}\relax
\EndOfBibitem
\bibitem[Soccodato \latin{et~al.}(2024)Soccodato, Penazzi, Pecchia, Phan, and Auf~der Maur]{MLApproach4}
Soccodato,~D.; Penazzi,~G.; Pecchia,~A.; Phan,~A.-L.; Auf~der Maur,~M. Machine learned environment-dependent corrections for a $spds^*$ empirical tight-binding basis. \emph{Mach. Learn.: Sci. Technol.} \textbf{2024}, \emph{5}\relax
\mciteBstWouldAddEndPuncttrue
\mciteSetBstMidEndSepPunct{\mcitedefaultmidpunct}
{\mcitedefaultendpunct}{\mcitedefaultseppunct}\relax
\EndOfBibitem
\bibitem[Gu \latin{et~al.}(2024)Gu, Zhou, Pandey, Zhang, Zhang, and E]{MLApproach5}
Gu,~Q.; Zhou,~A.; Pandey,~S.~K.; Zhang,~P.; Zhang,~L.; E,~W. Deep learning tight-binding approach for large-scale electronic simulations at finite temperatures with ab initio accuracy. \emph{Nat. Commun.} \textbf{2024}, \emph{15}\relax
\mciteBstWouldAddEndPuncttrue
\mciteSetBstMidEndSepPunct{\mcitedefaultmidpunct}
{\mcitedefaultendpunct}{\mcitedefaultseppunct}\relax
\EndOfBibitem
\bibitem[Kresse and Furthm\"uller(1996)Kresse, and Furthm\"uller]{Method_VASP}
Kresse,~G.; Furthm\"uller,~J. Efficiency of ab-initio total energy calculations for metals and semiconductors using a plane-wave basis set. \emph{Comput. Mater. Sci.} \textbf{1996}, \emph{6}, 15--50\relax
\mciteBstWouldAddEndPuncttrue
\mciteSetBstMidEndSepPunct{\mcitedefaultmidpunct}
{\mcitedefaultendpunct}{\mcitedefaultseppunct}\relax
\EndOfBibitem
\bibitem[Perdew \latin{et~al.}(1996)Perdew, Burke, and Ernzerhof]{Method_GGA}
Perdew,~J.~P.; Burke,~K.; Ernzerhof,~M. Generalized Gradient Approximation Made Simple. \emph{Phys. Rev. Lett.} \textbf{1996}, \emph{77}, 3865--3868\relax
\mciteBstWouldAddEndPuncttrue
\mciteSetBstMidEndSepPunct{\mcitedefaultmidpunct}
{\mcitedefaultendpunct}{\mcitedefaultseppunct}\relax
\EndOfBibitem
\bibitem[Perdew \latin{et~al.}(1998)Perdew, Burke, and Ernzerhof]{Method_PBE}
Perdew,~J.~P.; Burke,~K.; Ernzerhof,~M. Perdew, Burke, and Ernzerhof Reply:. \emph{Phys. Rev. Lett.} \textbf{1998}, \emph{80}, 891--891\relax
\mciteBstWouldAddEndPuncttrue
\mciteSetBstMidEndSepPunct{\mcitedefaultmidpunct}
{\mcitedefaultendpunct}{\mcitedefaultseppunct}\relax
\EndOfBibitem
\bibitem[Kresse and Furthm\"uller(1996)Kresse, and Furthm\"uller]{Method_planewavebasisset}
Kresse,~G.; Furthm\"uller,~J. Efficient iterative schemes for ab initio total-energy calculations using a plane-wave basis set. \emph{Phys. Rev. B} \textbf{1996}, \emph{54}, 11169--11186\relax
\mciteBstWouldAddEndPuncttrue
\mciteSetBstMidEndSepPunct{\mcitedefaultmidpunct}
{\mcitedefaultendpunct}{\mcitedefaultseppunct}\relax
\EndOfBibitem
\bibitem[Fuchs and Scheffler(1999)Fuchs, and Scheffler]{Method_PAW}
Fuchs,~M.; Scheffler,~M. Ab Initio Pseudopotentials for Electronic Structure Calculations of Poly-Atomic Systems Using Density-Functional Theory. \emph{Comput. Phys. Commun.} \textbf{1999}, \emph{119}, 67--98\relax
\mciteBstWouldAddEndPuncttrue
\mciteSetBstMidEndSepPunct{\mcitedefaultmidpunct}
{\mcitedefaultendpunct}{\mcitedefaultseppunct}\relax
\EndOfBibitem
\bibitem[Mostofi \latin{et~al.}(2008)Mostofi, Yates, Lee, Souza, Vanderbilt, and Marzari]{Method_Wannier901}
Mostofi,~A.~A.; Yates,~J.~R.; Lee,~Y.-S.; Souza,~I.; Vanderbilt,~D.; Marzari,~N. wannier90: A Tool for Obtaining Maximally Localised Wannier Functions. \emph{Comput. Phys. Commun} \textbf{2008}, \emph{178}, 685--699\relax
\mciteBstWouldAddEndPuncttrue
\mciteSetBstMidEndSepPunct{\mcitedefaultmidpunct}
{\mcitedefaultendpunct}{\mcitedefaultseppunct}\relax
\EndOfBibitem
\bibitem[Huang and Lee(2012)Huang, and Lee]{h-BN_NFE1}
Huang,~B.; Lee,~H. Defect and impurity properties of hexagonal boron nitride: A first-principles calculation. \emph{Phys. Rev. B} \textbf{2012}, \emph{86}, 245406\relax
\mciteBstWouldAddEndPuncttrue
\mciteSetBstMidEndSepPunct{\mcitedefaultmidpunct}
{\mcitedefaultendpunct}{\mcitedefaultseppunct}\relax
\EndOfBibitem
\bibitem[Hunt \latin{et~al.}(2020)Hunt, Monserrat, Z\'olyomi, and Drummond]{h-BN_NFE2}
Hunt,~R.~J.; Monserrat,~B.; Z\'olyomi,~V.; Drummond,~N.~D. Diffusion quantum Monte Carlo and $GW$ study of the electronic properties of monolayer and bulk hexagonal boron nitride. \emph{Phys. Rev. B} \textbf{2020}, \emph{101}, 205115\relax
\mciteBstWouldAddEndPuncttrue
\mciteSetBstMidEndSepPunct{\mcitedefaultmidpunct}
{\mcitedefaultendpunct}{\mcitedefaultseppunct}\relax
\EndOfBibitem
\bibitem[Marzari \latin{et~al.}(2012)Marzari, Mostofi, Yates, Souza, and Vanderbilt]{Method_Wannier902}
Marzari,~N.; Mostofi,~A.~A.; Yates,~J.~R.; Souza,~I.; Vanderbilt,~D. Maximally localized Wannier functions: Theory and applications. \emph{Rev. Mod. Phys.} \textbf{2012}, \emph{84}, 1419--1475\relax
\mciteBstWouldAddEndPuncttrue
\mciteSetBstMidEndSepPunct{\mcitedefaultmidpunct}
{\mcitedefaultendpunct}{\mcitedefaultseppunct}\relax
\EndOfBibitem
\bibitem[Dean~Moldovan(2020)]{Method_PyBinding}
Dean~Moldovan,~F.~P.,~Miša~Anđelković Pybinding v0.9.5: A Python package for tight-binding calculations. \emph{Zenodo} \textbf{2020}, \relax
\mciteBstWouldAddEndPunctfalse
\mciteSetBstMidEndSepPunct{\mcitedefaultmidpunct}
{}{\mcitedefaultseppunct}\relax
\EndOfBibitem
\bibitem[Wang \latin{et~al.}(2009)Wang, Lu, Yao, Li, Yip, and Ho]{TB_local1}
Wang,~C.-Z.; Lu,~W.-C.; Yao,~Y.-X.; Li,~J.; Yip,~S.; Ho,~K.-M. In \emph{Scientific Modeling and Simulations}; Yip,~S., de~la Rubia, Diaz,~T., Eds.; Springer Netherlands: Dordrecht, 2009; pp 81--95\relax
\mciteBstWouldAddEndPuncttrue
\mciteSetBstMidEndSepPunct{\mcitedefaultmidpunct}
{\mcitedefaultendpunct}{\mcitedefaultseppunct}\relax
\EndOfBibitem
\bibitem[Fang \latin{et~al.}(2015)Fang, Kuate~Defo, Shirodkar, Lieu, Tritsaris, and Kaxiras]{TB_local2}
Fang,~S.; Kuate~Defo,~R.; Shirodkar,~S.~N.; Lieu,~S.; Tritsaris,~G.~A.; Kaxiras,~E. Ab initio tight-binding Hamiltonian for transition metal dichalcogenides. \emph{Phys. Rev. B} \textbf{2015}, \emph{92}, 205108\relax
\mciteBstWouldAddEndPuncttrue
\mciteSetBstMidEndSepPunct{\mcitedefaultmidpunct}
{\mcitedefaultendpunct}{\mcitedefaultseppunct}\relax
\EndOfBibitem
\bibitem[Sovizi and Szoszkiewicz(2022)Sovizi, and Szoszkiewicz]{Case_study_1_TMDCs2}
Sovizi,~S.; Szoszkiewicz,~R. Single atom doping in 2D layered MoS2 from a periodic table perspective. \emph{Surf. Sci. Rep.} \textbf{2022}, \emph{77}, 100567\relax
\mciteBstWouldAddEndPuncttrue
\mciteSetBstMidEndSepPunct{\mcitedefaultmidpunct}
{\mcitedefaultendpunct}{\mcitedefaultseppunct}\relax
\EndOfBibitem
\bibitem[Jones \latin{et~al.}(2022)Jones, Xing, Swallow, Shiel, Featherstone, Smiles, Fleck, Thakur, Lee, Hardwick, Scanlon, Regoutz, Veal, and Dhanak]{Case_study_1_TMDCs1}
Jones,~L. A.~H.; Xing,~Z.; Swallow,~J. E.~N.; Shiel,~H.; Featherstone,~T.~J.; Smiles,~M.~J.; Fleck,~N.; Thakur,~P.~K.; Lee,~T.-L.; Hardwick,~L.~J. \latin{et~al.}  Band Alignments, Electronic Structure, and Core-Level Spectra of Bulk Molybdenum Dichalcogenides (MoS2, MoSe2, and MoTe2). \emph{J. Phys. Chem. C} \textbf{2022}, \emph{126}, 21022--21033\relax
\mciteBstWouldAddEndPuncttrue
\mciteSetBstMidEndSepPunct{\mcitedefaultmidpunct}
{\mcitedefaultendpunct}{\mcitedefaultseppunct}\relax
\EndOfBibitem
\end{mcitethebibliography}

\clearpage
\appendix
\setcounter{figure}{0}
\setcounter{table}{0}
\renewcommand{\thefigure}{\thesection\arabic{figure}}
\renewcommand{\thetable}{\thesection\Roman{table}}

\section{TB background and physical rationale of the unit-cell-based protocol}
\begin{table*}[ht]
    \centering
    \renewcommand{\arraystretch}{1}
    \setlength{\tabcolsep}{6pt}
    \begin{tabular}{cc|c|ccc}
    &      &Ce$_\text{Mo}$ MoS$_2$&MoS$_2$  &CeS$_2$  &CeS$_2$     \\
    &      &($5 \times 5 \times 1$) &(Relaxed)  &(Relaxed)  &(Lat Fixed)   \\
    \hline
    Lattice ($\text{\AA}$) &$\alpha$, $\beta$    &16.09      &3.18       &3.95       &3.18 \\
    \hline
    \multirow{2}{*}{$\theta$}   
        &S$^t$-Mo-S$^b$         &82         &81         &-          &-      \\
        &S$^t$-Ce-S$^b$         &73         &-          &69         &81     \\
    \multirow{2}{*}{$d^{TM-X}$}
        &Mo-S                       &2.41       &2.41       &-          &-      \\
        &Ce-S                       &2.66       &-          &2.76       &2.41   \\
    \hline
    \multirow{6}{*}{$\varepsilon^\mathrm{Mo}_{d}$} 
          &A$'_1$                 &-0.21      &-0.19      &-          &-      \\
          &($d_{z^{2}}$)          &           &(+0.02)    &           &       \\
          &E$'$                   &-0.05      &-0.06      &-          &-      \\
          &($d_{x^2-y^2},d_{xy}$) &           &(-0.01)    &           &       \\
          &E$''$                  &0.87       &0.94       &-          &-      \\
          &($d_{xz},d_{yz}$)      &           &(+0.07)    &           &       \\
    \multirow{6}{*}{$\varepsilon^\mathrm{Ce}_{d}$} 
          &A$'_1$                 &3.15       &-          &3.65       &3.73   \\
          &($d_{z^{2}}$)          &           &           &(+0.5)     &(+0.58)\\
          &E$'$                   &2.9        &-          &3.65       &3.82   \\
          &($d_{x^2-y^2},d_{xy}$) &           &           &(+0.75)    &(+0.92)\\
          &E$''$                  &4.49       &-          &4.55       &4.78   \\
          &($d_{xz},d_{yz}$)      &           &           &(+0.06)    &(+0.29)\\
    \multirow{6}{*}{$\varepsilon^\mathrm{Ce}_{f}$} 
          &A$'_1$                      &0.71  &-          &0.71       &1.72   \\
          &($f_{x(x^2-3y^2)}$)         &      &           &(+0)       &(+1.01)\\
          &A$'_2$                      &0.71  &-          &0.76       &1.6    \\
          &($f_{y(3x^2-y^2)}$)         &      &           &(+0.05)    &(+0.89)\\
          &A$''_2$                     &0.94  &-          &0.99       &1.93   \\
          &($f_{z^3}$)                 &      &           &(+0.05)    &(+0.99)\\
          &E$'$                        &0.9   &-          &1.01       &1.98   \\
          &($f_{yz^2}, f_{xz^2}$)      &      &           &(+0.11)    &(+1.08)\\
          &E$''$                       &0.9   &-          &0.94       &1.93   \\
          &($f_{xyz}, f_{z(x^2-y^2)}$) &      &           &(+0.04)    &(+0.93)\\
    \hline
    \multirow{6}{*}{$t^{\mathrm{Mo\text{--}X}}_{d\mu,j\nu}$}
        & \multirow{2}{*}{A$'_1$--A$_1$}
                                    &0.58       &0.6        &-          &-      \\
        &                           &           &(+0.02)    &           &       \\
        & \multirow{2}{*}{E$'$--E}
                                    &0.44       &0.44       &-          &-      \\
        &                           &           &(+0)       &           &       \\
        & \multirow{2}{*}{E$''$--E}
                                    &0.67       &0.65       &-          &-      \\
        &                           &           &(-0.02)    &           &       \\
    \multirow{6}{*}{$t^{\mathrm{Ce\text{--}X}}_{d\mu,j\nu}$}
        & \multirow{2}{*}{A$'_1$--A$_1$}
                                    &0.77       &-          &0.64       &0.63   \\
        &                           &           &           &(-0.13)    &(-0.14)\\
        & \multirow{2}{*}{E$'$--E}
                                    &0.49       &-          &0.37       &0.5    \\
        &                           &           &           &(-0.12)    &(+0.01)\\
        & \multirow{2}{*}{E$''$--E}
                                    &0.72       &-          &0.65       &0.85   \\
        &                           &           &           &(-0.07)    &(+0.13)\\
    \multirow{8}{*}{$t^{\mathrm{Ce\text{--}X}}_{f\mu,j\nu}$}
        & \multirow{2}{*}{A$'_2$--A$_1$}
                                    &0.15       &-          &0.16       &0.27   \\
        &                           &           &           &(+0.01)    &(+0.12)\\
        & \multirow{2}{*}{A$''_2$--A$_1$}
                                    &0.24       &-          &0.2        &0.39   \\
        &                           &           &           &(-0.04)    &(+0.15)\\
        & \multirow{2}{*}{E$'$--E}
                                    &0.14       &-          &0.11       &0.25   \\
        &                           &           &           &(-0.03)    &(+0.11)\\
        & \multirow{2}{*}{E$''$--E}
                                    &0.17       &-          &0.15       &0.22   \\
        &                           &           &           &(-0.02)    &(+0.05)\\
    \hline
    \end{tabular}       
    \caption{Comparison of TB parameters for Ce$_\mathrm{Mo}$ MoS$_2$ supercell and related unit cell: (i) a 5$\times$5$\times$1 Ce$_\mathrm{Mo}$ MoS$_2$ supercell with both lattice vectors and atomic positions fully relaxed, (ii) a host MoS$_2$ unit cell (fully relaxed), (iii) a CeS$_2$ unit cell (fully relaxed), and (iv) a CeS$_2$ unit cell with lattice vectors fixed to those of MoS$_2$ but atomic positions relaxed. Orbitals are grouped by local symmetry. On-site energies $\varepsilon^\mathrm{M}$ for Mo in the Ce$_\mathrm{Mo}$ MoS$_2$ supercell represent average values over the six Mo atoms surrounding the Ce substitution site. For Ce, both $d$- and $f$-orbital on-site energies are listed. Hopping amplitudes $t^{\mathrm{M\text{--}X}}$ between metal (M) and chalcogen (X) orbitals are averaged over all symmetry-equivalent bonds (e.g., for Ce--S, the absolute values of hopping elements between Ce and the six neighboring S atoms are averaged; the same applies to Mo--S). All hopping amplitudes are presented as absolute values. The values in parentheses indicate the deviation from the corresponding parameter in the Ce$_\mathrm{Mo}$ MoS$_2$ supercell. All energies are aligned with respect to the Fermi level of each system.}
    \label{SI_Table1}
\end{table*}

\begin{figure*}[ht]
    \centering
    \includegraphics[width=0.5\textwidth]{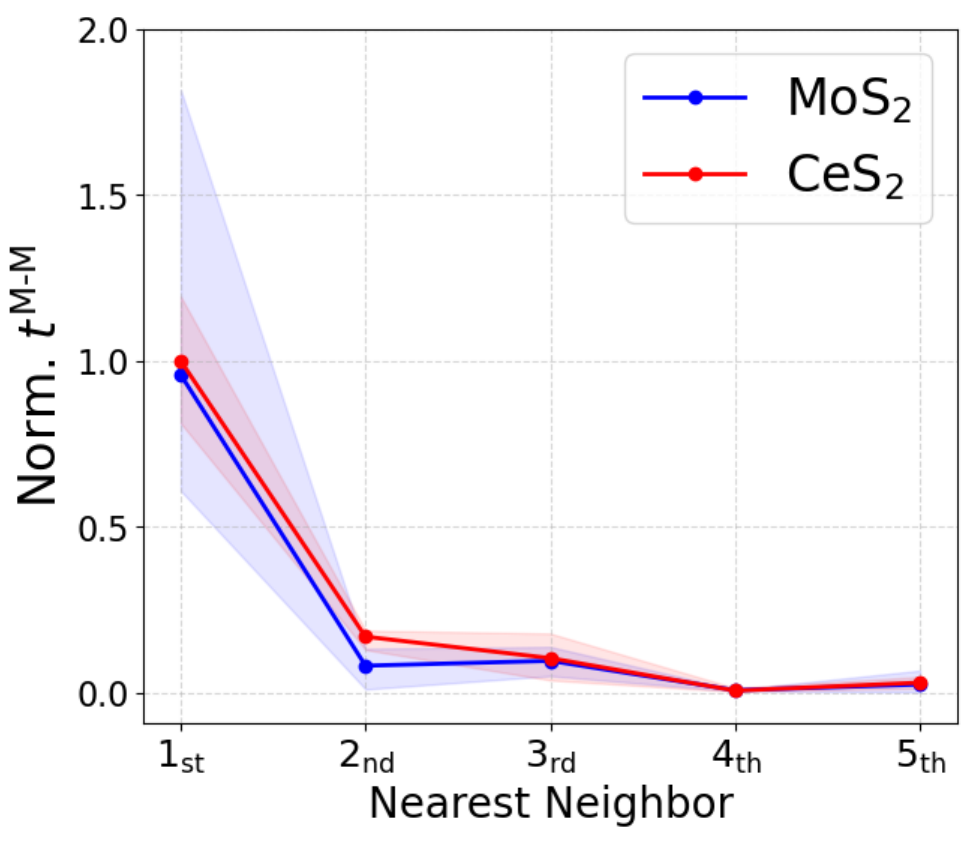}
    \caption{Decay of metal--metal hopping amplitudes with distance in MoS$_2$ and CeS$_2$ unit cells. Representative metal--metal hopping amplitudes $t^{\mathrm{M\text{--}M}}_{i\mu,j\mu}$ are plotted as a function of nearest-neighbor shell index for MoS$_2$ (blue) and CeS$_2$ (red). Each curve shows the mean value averaged over all symmetry-equivalent hopping elements within each shell, and the shaded region indicates the range spanned by the minimum and maximum values across different $d$-orbitals. All hopping amplitudes are normalized by the mean first nearest-neighbor hopping amplitude in CeS$_2$. The rapid decay—approximately an order of magnitude from the first to the second nearest-neighbor shell—demonstrates the local character of the TB interactions and provides the physical justification for restricting defect-site modifications to on-site energies and first-shell hopping parameters (if necessary) in the unit-cell-based protocol.}
    \label{SI_Fig_NN}
\end{figure*}

\section{Case study 3: NV\texorpdfstring{$^-$}{-} center in diamond}
\begin{figure*}[t]
    \centering
    \includegraphics[width=0.3\textwidth]{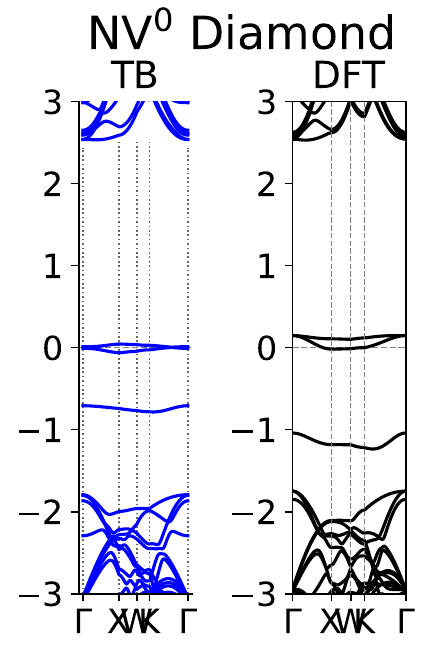}
    \caption{Band structures of NV$^0$ center diamond. The left (blue) bands are obtained from the unit-cell-based protocol, and the right (black) bands are calculated from DFT supercell calculations. The Fermi energy is set at $E_F = 0\ eV$.}
    \label{SI_Fig_Case3_BS}
\end{figure*}

\end{document}